%% file: PEEM_manuscriptv5.tex
\documentclass[aps,prb,twocolumn,10pt,showpacs,floatfix,superscriptaddress,longbibliography]{revtex4-1}

\usepackage{graphicx}
\graphicspath{{./figures/}}
\usepackage{color}
\usepackage{amsmath}
\usepackage{amssymb}
\usepackage{amsfonts}
\usepackage{upgreek}
\usepackage[utf8]{inputenc}

\begin{document}

\title{Antiferromagnetic-ferromagnetic phase domain development\\ in nanopatterned FeRh islands}

\author{R.~C.~Temple}
\email{r.c.temple@leeds.ac.uk}
\affiliation{School of Physics and Astronomy, University of Leeds, Leeds LS2 9JT, UK}

\author{T.~P.~Almeida}
\affiliation{School of Physics and Astronomy, University of Glasgow, Glasgow G12 8QQ, UK}

\author{J.~R.~Massey}
\affiliation{School of Physics and Astronomy, University of Leeds, Leeds LS2 9JT, UK}

\author{K.~Fallon}
\affiliation{School of Physics and Astronomy, University of Glasgow, Glasgow G12 8QQ, UK}

\author{R.~Lamb}
\affiliation{School of Physics and Astronomy, University of Glasgow, Glasgow G12 8QQ, UK}

\author{S.~A.~Morley}
\altaffiliation[Present address:]{Physics Department, University of California, Santa Cruz, CA 95064, USA}
\affiliation{School of Physics and Astronomy, University of Leeds, Leeds LS2 9JT, UK}

\author{F.~Maccherozzi}
\affiliation{Diamond Light Source Ltd, Harwell Science and Innovation Campus, Didcot OX11 0DE, UK}

\author{S.~S.~Dhesi}
\affiliation{Diamond Light Source Ltd, Harwell Science and Innovation Campus, Didcot OX11 0DE, UK}

\author{D.~McGrouther}
\affiliation{School of Physics and Astronomy, University of Glasgow, Glasgow G12 8QQ, UK}

\author{S.~McVitie}
\affiliation{School of Physics and Astronomy, University of Glasgow, Glasgow G12 8QQ, UK}

\author{T.~A.~Moore}
\affiliation{School of Physics and Astronomy, University of Leeds, Leeds LS2 9JT, UK}

\author{C.~H.~Marrows}
\email{c.h.marrows@leeds.ac.uk}
\affiliation{School of Physics and Astronomy, University of Leeds, Leeds LS2 9JT, UK}

\date{\today}

\begin{abstract}
The antiferromagnetic to ferromagnetic phase transition in B2-ordered FeRh is imaged in laterally confined nanopatterned islands using photoemission electron microscopy with x-ray magnetic circular dichroism contrast. The resulting magnetic images directly detail the progression in the shape and size of the FM phase domains during heating and cooling through the transition. In 5~$\upmu$m square islands this domain development during heating is shown to proceed in three distinct modes: nucleation, growth, and merging, each with subsequently greater energy costs. In 0.5~$\upmu$m islands, which are smaller than the typical final domain size, the growth mode is stunted and the transition temperature was found to be reduced by 20~K. The modification to the transition temperature is found by high resolution scanning transmission electron microscopy to be due to a 100~nm chemically disordered edge grain present as a result of ion implantation damage during the patterning. FeRh has unique possibilities for magnetic memory applications; the inevitable changes to its magnetic properties due to subtractive nanofabrication will need to be addressed in future work in order to progress from sheet films to suitable patterned devices.
\end{abstract}


\maketitle

\section{Introduction}

The binary equiatomic ordered alloy FeRh has a magnetostructural phase transition at an unusually high temperature of approximately 370~K \cite{Lewis2016}. This is a first order transition from an antiferromagnetic (AF) state to a ferromagnetic (FM) state and is accompanied by a 1\% uniform volume expansion of the crystal lattice. The transition temperature can be manipulated by magnetic field \cite{Maat2005}, strain \cite{Liu2016a,Cherifi2014}, chemical doping \cite{Barua2013b} and spin-polarised currents \cite{Suzuki2015}. Due to this high versatility this material has been extensively investigated in recent years for its potential technological applications in heat-assisted memory recording \cite{Thiele2003}, antiferromagnetic memory \cite{Marti2014, Moriyama2015} and electric field control of magnetic order \cite{Cherifi2014}.

The transition specifically occurs for the B2 ordering of the crystal lattice with the Fe atoms forming a simple cubic structure and Rh atoms at the body centre position. It is mainly driven by the free energy difference between the two electronic states of G-order AF on the Fe sites and an FM state with the Fe moments aligned with a 1~$\mu _\mathrm{B}$ moment that appears on the Rh site. Due to the first order nature of the transition, domains of both the AF and FM phases can coexist in the same sample \cite{Lewis2016}. Previous studies have imaged \cite{Baldasseroni2012, Kinane2014, Baldasseroni2015, Almeida2017a} and manipulated \cite{LeGraet2015} these coexisting states and nucleation and growth of FM regions in an AF matrix have been observed.

Future applications based on novel magnetic memory will require thin film material patterned on the nanoscale using known fabrication techniques. The properties of the FeRh phase transition are well known for bulk forms of FeRh but are modified with some variability when produced in a sheet film. FeRh films have been extensively studied and it is known that both strain and reduced symmetry at the interfaces combine to modify the transition temperature and broaden the transition width to various degrees throughout the depth of the film \cite{Ceballos2017,Gatel2017,Pressacco2016,Barton2017}. These modifications are dependent on film thickness, crystalline order and capping material.

Further changes to the transition arise when it is confined to lower dimensions through patterning, particularly due to phase domain shape and formation. Electronic transport in FeRh nanowires has been seen to show large supercooling effects and a highly asymmetric transition between heating and cooling \cite{Uhlir2016}. There have been few studies as yet on the effects of lateral confinement of the FeRh layer \cite{Arregi2018}.

In this investigation we use x-ray photo-emission microscopy (XPEEM) to obtain magnetically sensitive images of a patterned FeRh film. The lateral confinement is in square islands of side lengths 0.5, 1 and 5~$\upmu$m, below and above the final typical FM domain size of $\sim 1~\upmu$m \cite{Baldasseroni2012, Kinane2014, Baldasseroni2015, Almeida2017a}. Following a description of the fabrication and characterisation of our samples, our results are presented in two main sections. In the first section we examine the phase domain development in a 5~$\upmu$m island and show the three stages of domain creation and propagation. In the second section we compare this transition domain behaviour to the more confined 1 and 0.5~$\upmu$m islands and discuss the consequences of the changes observed in the context of future technological development.

\section{Material characterisation}

\begin{figure}
\centering
\includegraphics[width=8.5cm]{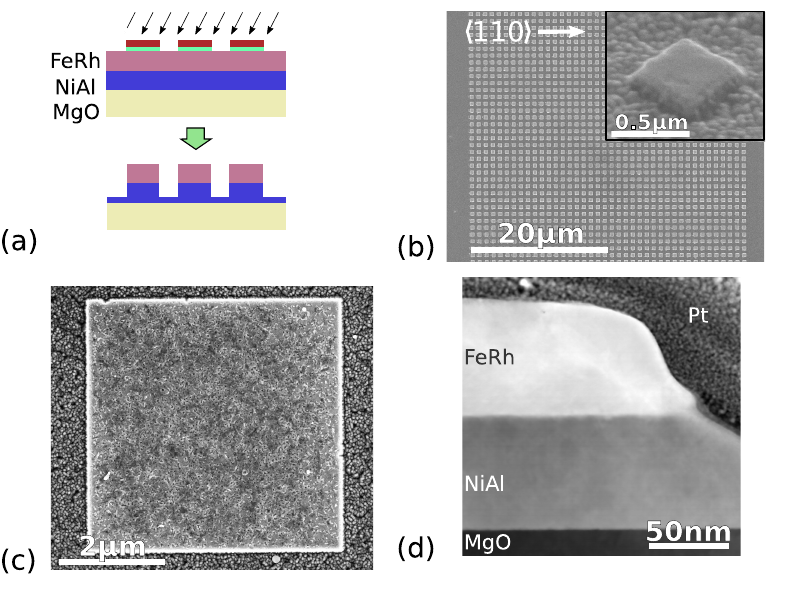}
\caption{FeRh nanopatterned islands. (a) Diagram of the nanofabrication process, see the text for full explanation. (b) SEM image of the array of 0.5~$\upmu$m squares. Inset is a close up image of one of the islands taken at a 45 degree angle to the film plane. The damage to the NiAl layer from the ion milling is seen. (c) In-lens detection SEM image of a 5~$\upmu$m island highlighting grain edges. (d) Cross-section TEM on the edge of a 5~$\upmu$m island using annular dark field contrast.}
\label{patterningfig}
\end{figure}

The patterned film that we studied is a NiAl(70~nm)/FeRh(70~nm) bilayer deposited on an MgO (001)-oriented substrate using DC magnetron sputtering at 600$^\circ$C, followed by a 60~min anneal at 700$^\circ$C. Further details of the FeRh deposition are given elsewhere \cite{LeGraet2013}. Standard e-beam lithography using a JEOL JBX-6300FS system followed by e-beam evaporation was used in a positive lift-off process to create an Al/Ti hard mask of square islands with side lengths of 0.5, 1 and 5~$\upmu$m, all separated on a grid structure with one island width gaps (the spacing was designed to negate magnetic dipolar interactions between the islands). The mask was transferred to the sheet film by Ar$^+$ ion milling from a 1~kV source at a 30$^\circ$ incidence angle under continuous rotation to a target thickness of 20~nm NiAl. The NiAl is a paramagnetic metal with a B2 structure lattice matched to the FeRh and is required as a path to ground for the photo-emission current during the XPEEM experiment. Finally the metal mask was removed with a standard Al chemical etch to leave the reveal the bare FeRh surface. A diagram of this process is shown in Fig.~\ref{patterningfig}(a). All of the islands shown in this paper were created with straight edges along the $\langle 110\rangle$ crystal orientations of the FeRh. XPEEM is a surface sensitive technique so the chemical etch was employed as it was necessary to protect the FeRh surface from the ion milling. FeRh is resilient to chemical degradation and oxidation due to the high Rh content \cite{Baldasseroni2014}.

Fig.~\ref{patterningfig}(b) shows a scanning electron microscope (SEM) secondary electron emission image of a final patterned array of NiAl/FeRh islands of side 0.5~$\upmu$m. In the inset shown from a 45$^\circ$ angle to the film surface the rough surface of the surrounding NiAl is seen. This is thought to be due to preferential milling at grain boundaries enhancing any roughness present in the as-deposited film. This is also seen in Fig.~\ref{patterningfig}(c), an SEM image of a 5~$\upmu$m island using an in-lens detector to highlight gradients in the topography.

Scanning transmission electron microscopy (STEM) was utilised to image a cross-section of the sample taken through a 5~$\upmu$m island. The sectioned specimen was created using a focussed ion beam ``liftout" technique \cite{Almeida2017a} and produced with an estimated thickness between 50 and 120~nm. Fig.~\ref{patterningfig}(d) shows an annular dark field image, sensitive to atomic number contrast, obtained from one of the edges of the island. The MgO, NiAl and FeRh layers are easily distinguished, as are the protective electron and ion beam deposited Pt layers from the cross-sectioning process. The fabricated island is observed to have a sloping edge with average angle of 30 degrees relative to the vertical.

\begin{figure}
\centering
\includegraphics[width=7.5cm]{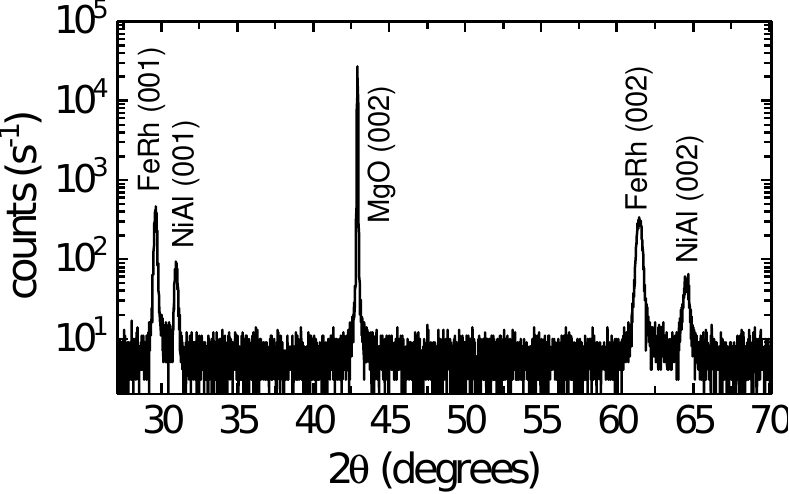}
\caption{X-ray diffraction of the NiAl/FeRh sheet film in the Bragg-Brentano geometry showing the B2 crystal structure. }
\label{xrayfig}
\end{figure}

The epitaxy of the material was confirmed with x-ray diffraction shown in Fig.~\ref{xrayfig}. The B2 (CsCl) phase of the FeRh is identified with the clear (001) superstructure peak; no other phases are detected. The ratio of the (001) to (002) peak intensities gives a chemical order parameter of 0.8 (where 1 is fully B2 ordered and 0 is in the random bcc state \cite{deVries2013}). The NiAl is also in the B2 phase with a lattice parameter of 2.884~\AA. The bulk lattice parameter of FeRh is 2.99~\AA \cite{Ceballos2017} so that the NiAl introduces lattice mismatch strain of -5.3\% on the FeRh film. The FeRh lattice parameters of the free film were determined to be $a=2.972$~\AA\ and $c=3.013$~\AA\ showing the expected tetragonal distortion.

The XPEEM imaging was done on beamline I06 at Diamond Light Source in the XPEEM endstation on and off the Fe L$_\mathrm{3}$ resonance. The probing depth of the technique is approximately 3~nm. Electron yield intensity images were recorded using positive circular polarised $I^+$ and negative circular polarised radiation $I^-$ on (at energy 706.6~eV) and off (700~eV) the Fe L$_\mathrm{3}$ absorption edge. On the L$_\mathrm{3}$ edge there will be x-ray magnetic circular dichroism (XMCD) contrast that yields information about FM ordered regions. The four sets of images of each polarisation on and off the resonance edge were averaged and then normalised to a defocussed image and the off resonance images were subtracted to obtain the difference signal $d^{\pm}=I_{on}^{\pm}-I_{off}^{\pm}$. The final quantity displayed in the XMCD images is given as $D = (d^{+}-d^{-})/(d^{+}+d^{-})$.

\section{Results}
\subsection{Magnetic and phase domain behaviour}

\begin{figure*}
\centering
\includegraphics[width=15cm]{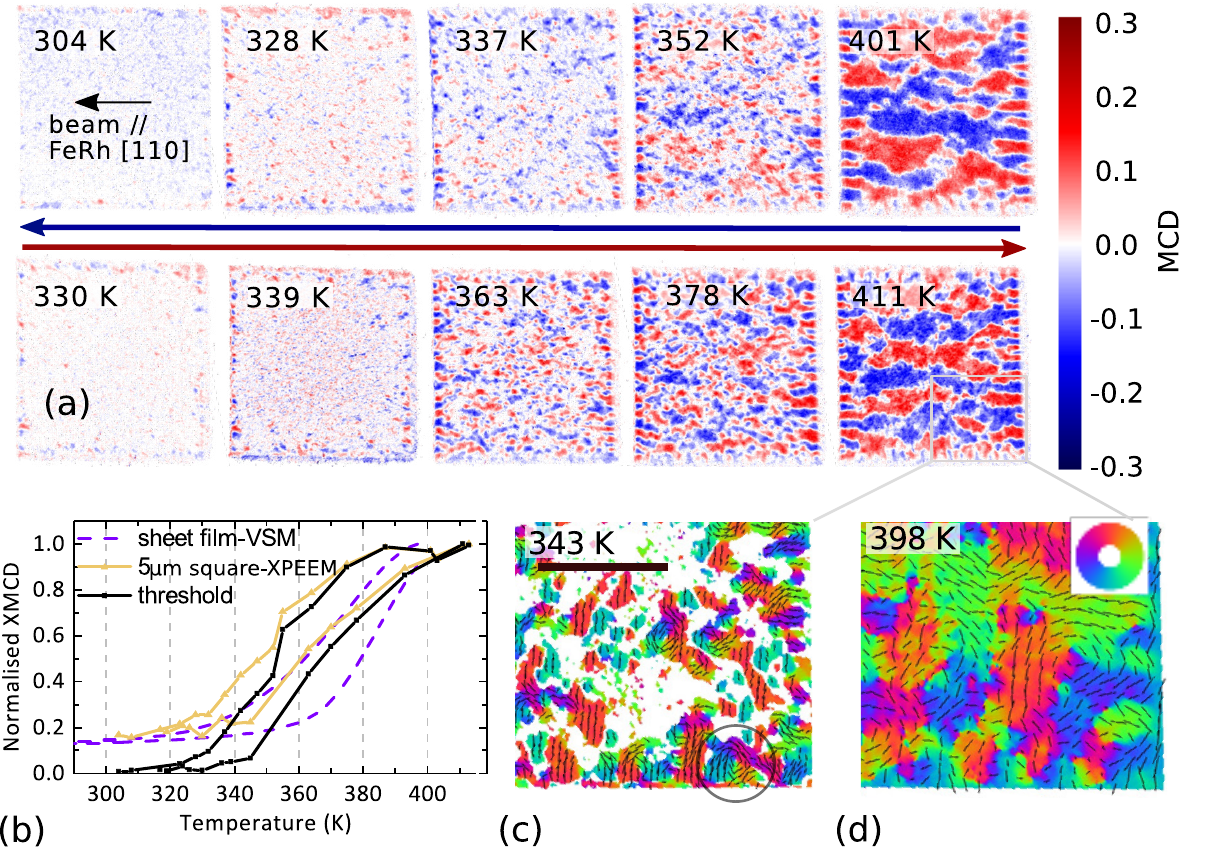}
\caption{XPEEM imaging of FM domains. (a) Magnetic dichroism images of an island of side 5~$\upmu$m shown as a function of temperature during the cooling process (upper row of images) and the heating process (lower row). The beam direction and magnetic sensitivity are shown in the top left image. (b) The integrated absolute XMCD signal as a function of temperature and in comparison to a VSM measurement. Both signals are normalised to the saturation value. The thresholded signal is also shown, see text for details. (c) and (d) Vector magnetisation maps of the bottom right portion of the 5~$\upmu$m island during heating. The inset scale bar on (c) is 1~$\upmu$m. The colour represents the vector direction indicated in the colour wheel and the length of the arrows represent local XMCD magnitude. A vortex state is highlighted in (c).}
\label{5um-mcdfig}
\end{figure*}

In Fig.~\ref{5um-mcdfig}(a) the dichroism images from a 5~$\upmu$m island are shown at a series of temperatures across the transition. The color map shows red and blue domains that are ferromagnetic with magnetisations having components parallel and antiparallel to the incoming beam direction. White areas have no XMCD signal and are either antiferromagnetic or magnetised perpendicular to the beam direction. As the temperature increases the ferromagnetic regions grow until the entire square has reached the ferromagnetic phase, and has a central magnetic domain width on the scale of 1~$\upmu$m. As has been previously observed in sheet film samples \cite{Baldasseroni2012}, the reverse transition occurs in a qualitatively similar manner.

The summed absolute XMCD signals for images at each temperature are plotted in orange in Fig.~\ref{5um-mcdfig}(b) and compared to SQUID vibrating sample magnetometry (VSM) measurements of the complete film before patterning. Both signals are normalised relative to their maximum value and plotted on the same scale. The VSM measurements were done under a saturating field of 0.1~T while the XPEEM measurements are necessarily done in a remanent state. An approximately 10~K hysteresis is seen in the transition on heating and cooling for each measurement. This is a well documented effect in the AF-FM FeRh and other first order transition systems where phases can coexist in the same volume \cite{Lewis2016}. While the general shape of the loop is the same in both measurements there is a clear 20~K offset between the VSM and XPEEM data. This is thought to be due to the difference between the VSM technique which is detecting the average magnetisation throughout the sheet film while the XPEEM is only looking at the surface state in a small region of the film. Thin film FeRh has been shown to have a lower transition temperature by up to 70~K at the top surface particularly in capped samples, but this has been observed even when uncapped \cite{Kim2009, Gatel2017, Fan2010}. This is generally attributed to strain relaxation and symmetry breaking at the interface. We cannot rule out some contribution of the change in transition temperature resulting from uncertainties in the thermometry from the two techniques also.

The non-zero magnetisation in the AF state seen in the VSM data is a common effect in FeRh and studies have shown this is due to a small stabilised ferromagnetic region close to the bottom interface of the film \cite{Pressacco2016,Gatel2017,Fan2010}. The non-zero signal we see in the XPEEM data at low temperatures however is caused by residual noise in regions of the island with zero magnetisation, this does not average to zero since we take the integral over the absolute XMCD signal $D$. A small threshold of $|D|>0.03$ is applied to the signal to eliminate this noise prior to the integral (the root mean square amplitude of the noise is 0.01). Using this threshold produces the black curve shown in Fig.~\ref{5um-mcdfig}(b). Within the experimental limitation no magnetic signal is detected in AF state at the top surface.

The magnetic domains are examined in Fig.~\ref{5um-mcdfig}(c) and (d). These are vector maps, obtained by combining dichroism images at two perpendicular rotations of the sample with respect to the x-ray beam. At mid-transition point these domains form flux closure structures with a vortex state particularly visible in Fig.~\ref{5um-mcdfig}(c). In the fully ferromagnetic state the flux closure still exists but much larger domains are present.

While the dichroism images contain information on the direction of the magnetisation it is important to distinguish between standard magnetic domains - ubiquitous in ferromagnetic materials - and \emph{phase} domains which are contiguous volumes that have undergone transition to the FM state regardless of their magnetic direction.  These phase domains can be best seen using the absolute dichroism signal $|D|$ which makes no distinction between parallel and antiparallel magnetisations.

\begin{figure}
\centering
\includegraphics[width=6cm]{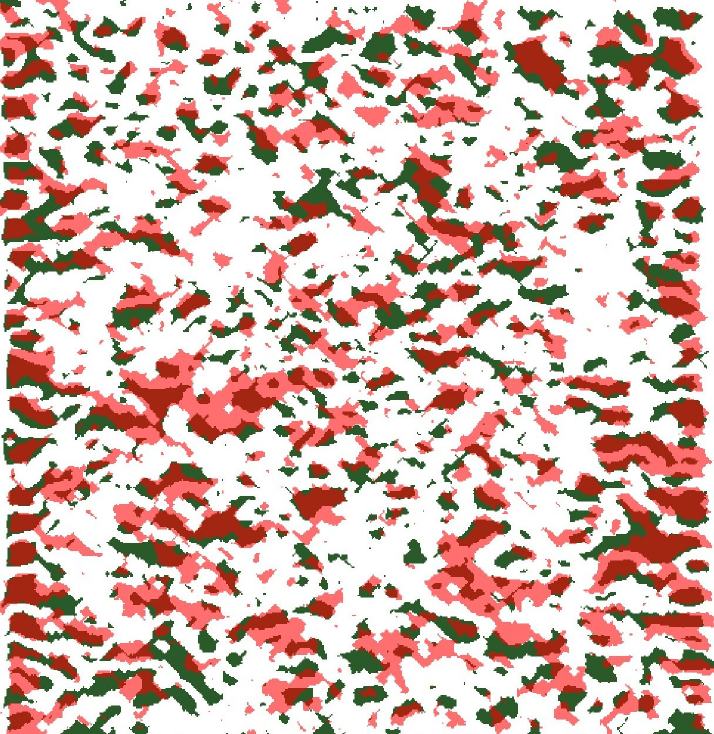}
\caption{The 363~K XMCD image from the heating arm (green) and 352~K XMCD image from the cooling arm (light red) made binary using a threshold value and overlaid for comparison (overlap dark red). The full 5~$\upmu$m island is shown. A high percentage of the domain sites are seen to overlap.}
\label{overlayfig}
\end{figure}

From the images in Fig.~\ref{5um-mcdfig}(a) we see the phase domain creation proceeds by a nucleation and growth mechanism. Looking in more detail at the heating and cooling arms many of the local area transition temperatures are seen to be similar in both arms for comparable states. In Fig.~\ref{overlayfig} the image at 363~K on the cooling arm has been overlaid on the image at 352~K on the heating arm. These are both shown as binary images so that all ferromagnetic domains appear one colour and areas with no magnetic signal are white. 44\% of the FM pixels in the heating arm overlap with the cooling arm as opposed to an expected 26\% for a random distribution of the pixels in the image.

One of the complications in applying simple first order models to FeRh is that the volume expansion of the lattice as it undergoes transition will create a dynamical strain landscape over the material as phase domains alter. This local strain may affect the development of local domains, allowing them to grow at lower temperatures than they would otherwise be able. Tensile strain is known to lower the transition temperature \cite{Liu2016a}.  In this data we see no obvious preference for nucleation of domains close to other current nucleated sites. It is clear from the similar domain positions in the overlaid images however, that there is a static background energy landscape that plays a larger role in the transition than the dynamic landscape, at least on the $>$100~nm length and 30~s time scale.

\begin{figure}
\centering
\includegraphics[width=6cm]{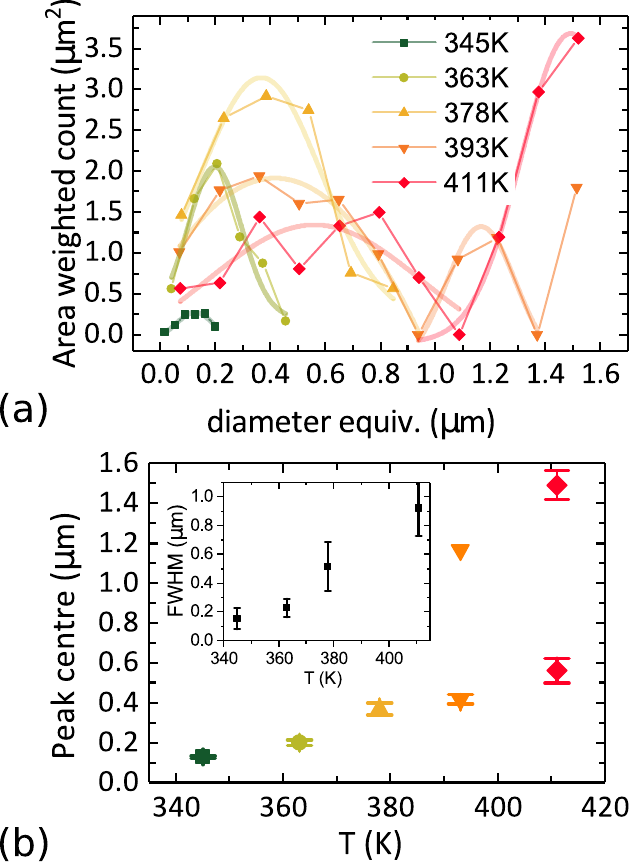}
\caption{Characteristic sizes of FM domains. (a) Area weighted histogram of the domain size at various temperatures during the heating process. See text for details. (b) Peak centres of Gaussian peak fits to the histograms in (a), inset the peak full width half maxima (FWHM) are plotted.}
\label{domainsizefig}
\end{figure}

We can further examine the phase domain formation using the binary images to find the sizes of the phase domains. In Fig.~\ref{domainsizefig}(a) an area weighted histogram of the domain equivalent diameter $d_\mathrm{eq}$ is plotted for various temperatures on the heating arm. The count has been weighted by the domain area $A$ in order to balance the space preference for large compared to small domains. The domain characteristic size $d_\mathrm{eq}$ is calculated simply from the domain area as $d_\mathrm{eq}=2\sqrt{A/\pi }$.

From 345 to 378~K the weighted histograms are simple peaked distributions. As the temperature increases the total number of domains increases as does the peak centre size and the peak width. The histogram peaks have all been fitted with Gaussian distributions which are shown on the histogram and the peak centres and widths are plotted in Fig.~\ref{domainsizefig}(b). Importantly this demonstrates that both nucleation and growth are taking place simultaneously at these lower temperatures. Domains nucleate at a typical size of $<$200~nm. From the histogram we see the majority of the domains are newly nucleated below 363~K. From 363 to 378~K the majority of the phase domain area increase proceeds by domain growth as opposed to nucleation as can be visibly seen in the dichroism images. At 393~K there remains a broad low diameter peak but a new region of the distribution opens up at larger diameters greater than 1~$\upmu$m. This is the final stage of phase growth where percolation has been reached and domains coalesce with neighbouring domains. Finally at 411~K the majority of the central domains are merged and form a peak in this larger diameter region. The edge domains remain somewhat separated and smaller than the central domain.

These histograms make the boundary between the three growth modes very clear in a way not observed before. It is only when there is little growth available left to the FM domains that they will merge. There is clearly an energy cost in the magnetostatic energy of the domain formation. It may be due to the topological stability of the vortex cores in the smaller domains.

It is noted that the final magnetic domains have an elongated shape along the beam direction in this case. This nematic ordering has been seen before in XPEEM experiments on Pd doped FeRh and was attributed to a martensitic transition \cite{Kinane2014}. In this case without the Pd doping that cannot be the case. FeRh has a very weak (cubic) anisotropy \cite{Mariager2013} and small stray magnetic fields can generate a stripe domain magnetostatic locking pattern \cite{Cohen1962}. This could potentially be the reason for the nematic ordering, however further investigations under a controlled magnetic field are necessary.

\subsection{Size effects}

\begin{figure}
\centering
\includegraphics[width=7cm]{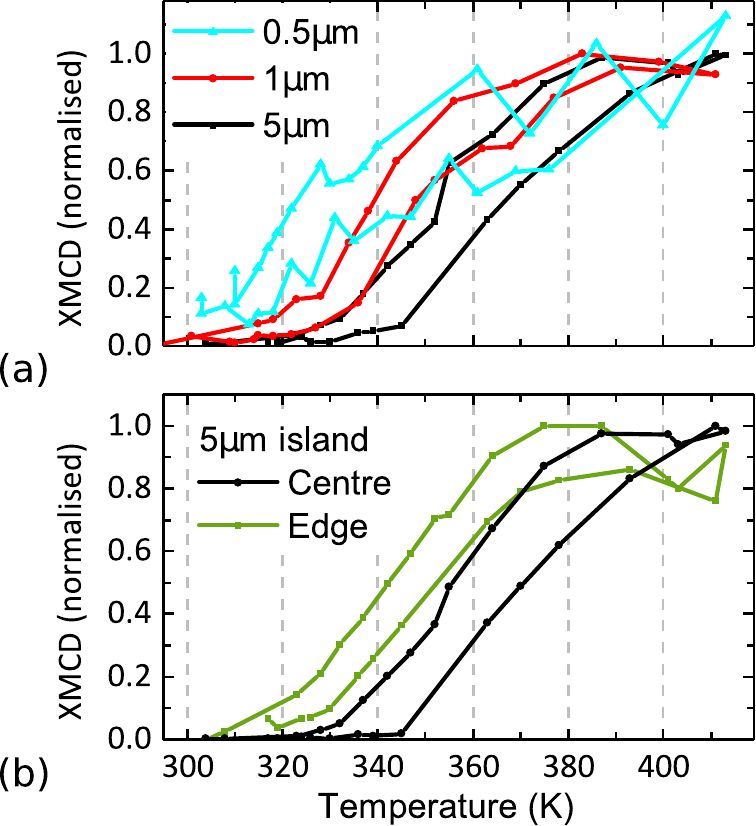}
\caption{Phase transition measured by XMCD. (a) Absolute dichroism integrated over the entire island for 0.5, 1 and 5~$\upmu$m island sizes shown on the same scale. For the smaller two sizes an average signal over four FeRh islands have been plotted. (b) The integrated dichroism signal vs.\ temperature for the 5um island. The integral is over the edge and central area of the island as indicated. }
\label{hysteresisfig}
\end{figure}

The integrated signal plot for the 0.5, 1 and 5~$\upmu$m island sizes is shown in Fig.~\ref{hysteresisfig}(a). As discussed before, asymmetry levels below 0.03 have not been included in the integration and the signals have been normalized to their saturation value. The transition temperature is shown to be reduced in the smaller islands. The transition temperature is calculated at the midpoint of the transition where 50\% of the magnetic signal remains. On the cooling arm it is found that the 1~$\upmu$m size has a transition temperature of 343~K, 12~K cooler than the 5~$\upmu$m island, and the 0.5~$\upmu$m island size $\approx 333$~K, a further $10$~K below that. The smaller islands show significantly more noise in the integrated signal making exact quantification of this offset difficult. This noise is a systematic issue caused by the sensitivity of the smaller images to the background subtraction and drift correction process.

\begin{figure}
\centering
\includegraphics[width=7cm]{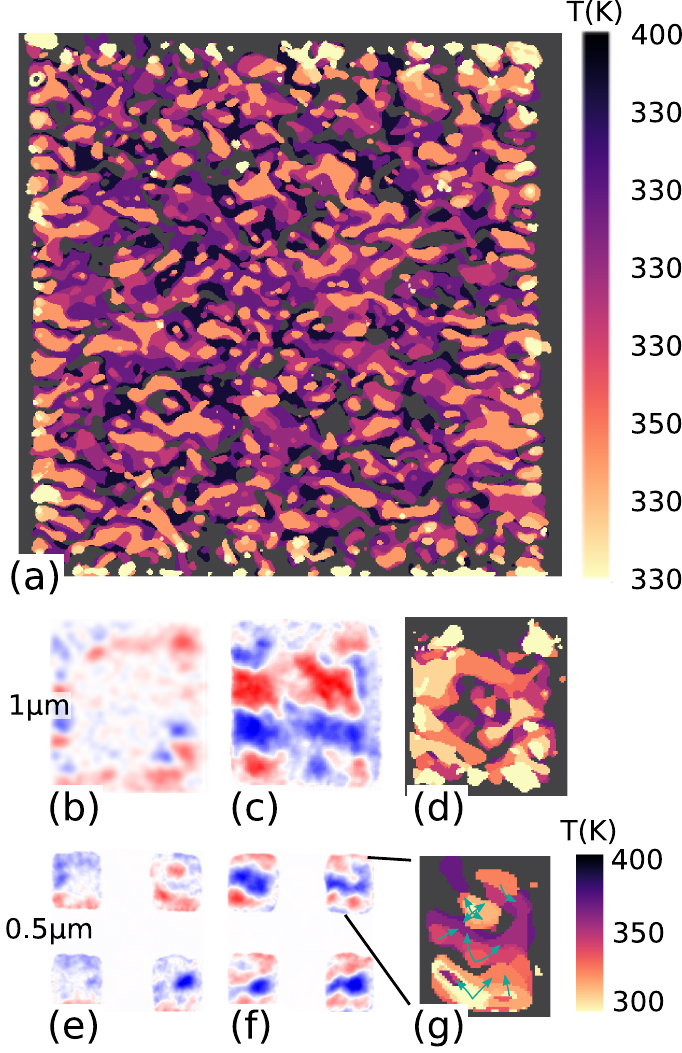}
\caption{Size dependence of the phase transition temperature. (a) Map of the phase transition temperature for a 5~$\upmu$m island. The absolute signal is used so only phase domains are shown. (b) and (c) show dichroism images from a 1~$\upmu$m island using the same color scheme as Fig. \ref{5um-mcdfig} at 336 and 411~K respectively and (d) is the temperature map of the transition using the same colour scale as for (a). (e) and (f) are dichroism images for the 500nm island size with edge effects less evident. (g) is the temperature map of the highlighted island in (f) with a slightly altered colour scale shown. Domain nucleation sites are isolated bright regions; on map (g) the domain propagation directions from the nucleation sites are indicated by arrows.}
\label{tempmapfig}
\end{figure}

It is clear from the images of the 5~$\upmu$m island in Fig.~\ref{5um-mcdfig}(a) that the transition temperature is reduced at the edge of the island. Phase domain nucleation occurs at a lower temperature on the edges of the patterned island and these domains grow while the central domains are still nucleating. The hysteresis loops of the edge and central region are compared directly in Fig.~\ref{hysteresisfig}(b). At 339~K during the heating process the majority of the island within 300~nm of the edge has nucleated ferromagnetic states while the middle of the square has approximately 9 nucleated domains comprising approximately 0.01\% of the central island area. In Fig.~\ref{tempmapfig}(a) we show a colour coded map of the transition temperature over the image area. For this image and other transition maps in the figure a 6 pixel moving average smoothing has been applied to the raw data, and the threshold for transition is set at an asymmetry of $|D|>0.015$. The edge nucleations are highlighted by the brighter colours around the edge and nucleated areas are seen to grow inwards towards the centre of the island as the temperature increases.

These edge effects are expected to play a more important role for smaller islands, and indeed this is what we find. Fig.~\ref{tempmapfig}(b) and (c) show dichroism images of a 1~$\upmu$m island during nucleation at 336~K and post-transition at 411~K respectively. We see in Fig.~\ref{tempmapfig}(b) the same edge state nucleation pattern. In this case the edge width is 200~nm. Again, this is highlighted in the transition map shown in Fig.~\ref{tempmapfig}(d), where the growth inwards of the edge nucleations is clearly seen.

For 0.5~$\upmu$m islands, which were the smallest we could obtain a clear dichroism signal from, a separate edge nucleation process is not evident. There are fewer nucleation events so the statistics are poor but these are not seen to be biased towards the edge regions of the island. Images including four such adjacent islands are shown in Fig.~\ref{tempmapfig}(e) and (f) during the nucleation phase, and post-transition. A transition temperature map of the upper right island is shown in Fig.~\ref{tempmapfig}(g). In the map we see that just six nucleation events trigger the phase domain growth over the island, propagation vectors for this growth are shown. This is perhaps not unexpected, since the edge domains on the larger islands were $>$200~nm. For a 500~nm island nearly the entire surface will be within the effective edge region.

The initial edge nucleation in the 5 and 1~$\upmu$m islands occur at similar temperatures of 330~K. The offset in the average transition temperature is simply due to the relative proportion of the island in the edge nucleation region: 19\% and 64\% of the total island area respectively for the two sizes. The initial nucleation of the 0.5~$\upmu$m island is at approximately 310~K, lower than for the larger islands. The offset in the phase transition temperature at the higher temperatures is due to the rapid progress of the transition once nucleated. Fewer nucleation events mean that fewer phase domain merging energy barriers have to be overcome, as with the edge of larger islands the FM domains rapidly grow to fill the space.

The question which remains is, what is the mechanism by which the edge properties are modified? There are two likely potential causes. Subtractive patterning of the sheet FeRh film, forming edges (of the islands), enables the release of strain caused by mismatch of the natural FeRh lattice parameter with the MgO substrate lattice parameter. In work on other materials this has been found to relieve up to 50\% of the strain in 500~nm pillars \cite{Himcinschi2007}. This alone would certainly generate a large change in the transition parameter; a change of 30~K has been measured previously for thick FeRh films under differing strain conditions \cite{Ceballos2017}.

A second possible cause is collision damage from the Ar$^{+}$ ion milling process. Ion milling damage during the mask transfer is a well documented phenomenon in magnetic media patterning \cite{Shaw2008} and other fields. Due to the titanium mask all damage to the FeRh island will be at the side-walls. FeRh is known to be particularly sensitive to ion damage \cite{Cervera2017} and its signature has been observed in TEM cross-sectional samples of FeRh prepared by 2keV Ar$^{+}$ ion milling \cite{McLarenThesis}. Controlled ion beam damage experiments have shown that low doses will initially lower the transition temperature by several tens of Kelvin followed by complete removal of the transition at higher doses, leaving a purely ferromagnetic material \cite{Cervera2017}. Using TRIM simulation \cite{TRIMsoftware} for 1~keV Ar$^{+}$ ions impinging on FeRh we find a penetration depth of up to 5~nm. However this does not take into account ion-channelling effects in crystalline materials which can significantly increase this distance \cite{Dobrev1982}. 

\begin{figure*}
\centering
\includegraphics[width=15cm]{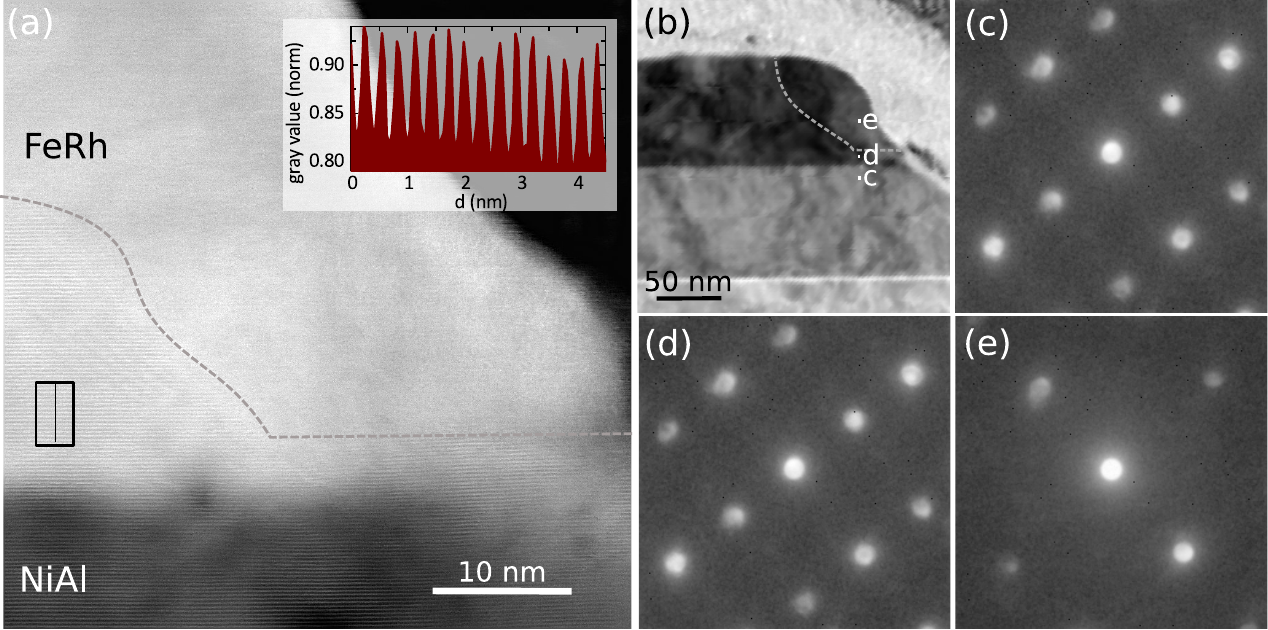}
\caption{STEM analysis of the patterned island edge. (a) High resolution STEM image of the right hand edge of the 5~$\upmu$m island shown in Fig.~\ref{patterningfig}(d). The inset is a vertical line scan over the indicated region showing lattice fringes. These fringes are not observed in the edge grain region above and right of the dashed grey line. (b) Bright field image with diffraction contrast over a larger region and grey dashed line showing continuation of grain. Nanodiffraction images of points indicated in (b) are shown in (c)-(e).}
\label{hrtemfig}
\end{figure*}

In order to investigate further the cause of reduced transition temperatures in the vicinity of the edges, high resolution (HR) STEM imaging was performed. An HRSTEM image of the edge region is shown in Fig.~\ref{hrtemfig}(a). Contrast in the image is dominated by atomic number and hence the FeRh layer exhibits higher average intensity than the NiAl buffer layer. In a region towards the left hand side of the FeRh layer, greater than $\sim30$~nm away from the island edge, and throughout the NiAl layer, horizontally oriented lattice fringes are observed. In the FeRh layer the periodicity of these fringes is $0.294\pm0.004$~nm, consistent with the FeRh lattice parameter and thus relates to the $\langle$001$\rangle$ lattice planes. Such fringes are only allowed to be visible where the B2, chemically ordered crystal structure exists (they are forbidden for chemically disordered bcc). Hence, the complete presence of fringes in the NiAl layer but only in part of the FeRh located away from the edge is suggestive of patterning induced structural modification.

The nature of the crystal structure modification in this region was further investigated using 4D nano-beam diffraction imaging. The edge region was probed with a parallel ($<$1 milliradian semi-convergence angle), nanofocused electron beam (6~nm diameter). Scanning the beam, diffraction patterns were recorded using a high speed electron counting detector \cite{QuantumDetectors} at each point in a $256 \times 256$ raster. Figures \ref{hrtemfig}(b-e) show this analysis, highlighting the image (essentially a bright field image showing diffraction contrast) and diffraction patterns from the regions where lattice fringes were observed in the NiAl and FeRh and from the edge region where fringes were absent in Fig.~\ref{hrtemfig}(a).

In Fig.~\ref{hrtemfig}(c) and (d) the diffraction patterns from the FeRh and NiAl are identical in form and consistent with diffraction from the B2 structure when viewed along the $[011]$ direction. In the region nearest the edge, figure \ref{hrtemfig}(e), the diffraction pattern contains significantly fewer spots and is consistent with a chemically disordered fcc crystal structure in this region. Relating this observation to the island edge contrast apparent in Fig.~\ref{hrtemfig}(b) (and also more weakly in the HAADF image shown in Fig.~\ref{patterningfig}(d)), it can be seen that the structural modification extends over a larger region, composing a ``grain'' which extends to approximately 100~nm in from the island edge and up to the top surface of the FeRh. Similar image contrast was observed at the opposite edge of the island (not shown).

While this chemically disordered modified crystal structure region is smaller than the approximately 250~nm edge domain nucleation region seen in the PEEM images it is likely to be the cause of the lowered transition temperature due the high sensitivity of the transition on the alloy chemical order and structure \cite{Cervera2017}. The mismatch between the size of the observed edge ``grain'' and the nucleated domains may simply be a consequence of where the cross-section was taken, the disordered grain may vary in width along the length of the island as the magnetic domains do.

\section{Conclusion}

The mixed phase domains in FeRh have been imaged during transition and have shown three distinct modes of development: nucleation, growth, and merging each proceeding at a higher energy cost than the last. The cost of the latter mode is attributed to the magnetostatic energy of ferromagnetic domain formation. The mixed phase domains have been studied in confined FeRh structures with sizes above and below the phase domain length scale. For 5 and 1~$\upmu$m islands the edge and central regions were separated with a 14~K reduction in the transition temperature on the edges. The cause of this reduction in transition temperature is a chemically disordered region within 100~nm of the island edge observed in cross-sectional HRSTEM. This is thought to arise from collision damage during low energy ion milling. For the 0.5~$\upmu$m structure the transition temperature was lowered a further 10~K due to a reduced number of merging events in the smaller space. In general as patterning is used to produce features of ever smaller size particular care will need to be taken over inevitable side wall damage in subtractive patterning for which FeRh is particularly sensitive. Strain relief is inevitable but its effects would be reduced in thinner films.

\begin{acknowledgments}
This work was financially supported by EPSRC under grant number EP/M018504/1 and EP/M019020/1. We thank Diamond Light Source for beamtime and we are grateful to M. C. Rosamond for assistance in the nanofabrication process and G. Stefanou for work on the SEM imaging.
\end{acknowledgments}

\input{finalbbl.bbl}
\end{document}

%% file: finalbbl.bbl
%

%% file: PEEM_manuscriptv5.bbl
\begin{thebibliography}{34}%
\makeatletter
\providecommand \@ifxundefined [1]{%
 \@ifx{#1\undefined}
}%
\providecommand \@ifnum [1]{%
 \ifnum #1\expandafter \@firstoftwo
 \else \expandafter \@secondoftwo
 \fi
}%
\providecommand \@ifx [1]{%
 \ifx #1\expandafter \@firstoftwo
 \else \expandafter \@secondoftwo
 \fi
}%
\providecommand \natexlab [1]{#1}%
\providecommand \enquote  [1]{``#1''}%
\providecommand \bibnamefont  [1]{#1}%
\providecommand \bibfnamefont [1]{#1}%
\providecommand \citenamefont [1]{#1}%
\providecommand \href@noop [0]{\@secondoftwo}%
\providecommand \href [0]{\begingroup \@sanitize@url \@href}%
\providecommand \@href[1]{\@@startlink{#1}\@@href}%
\providecommand \@@href[1]{\endgroup#1\@@endlink}%
\providecommand \@sanitize@url [0]{\catcode `\\12\catcode `\$12\catcode
  `\&12\catcode `\#12\catcode `\^12\catcode `\_12\catcode `\%12\relax}%
\providecommand \@@startlink[1]{}%
\providecommand \@@endlink[0]{}%
\providecommand \url  [0]{\begingroup\@sanitize@url \@url }%
\providecommand \@url [1]{\endgroup\@href {#1}{\urlprefix }}%
\providecommand \urlprefix  [0]{URL }%
\providecommand \Eprint [0]{\href }%
\providecommand \doibase [0]{http://dx.doi.org/}%
\providecommand \selectlanguage [0]{\@gobble}%
\providecommand \bibinfo  [0]{\@secondoftwo}%
\providecommand \bibfield  [0]{\@secondoftwo}%
\providecommand \translation [1]{[#1]}%
\providecommand \BibitemOpen [0]{}%
\providecommand \bibitemStop [0]{}%
\providecommand \bibitemNoStop [0]{.\EOS\space}%
\providecommand \EOS [0]{\spacefactor3000\relax}%
\providecommand \BibitemShut  [1]{\csname bibitem#1\endcsname}%
\let\auto@bib@innerbib\@empty
\bibitem [{\citenamefont {Lewis}\ \emph {et~al.}(2016)\citenamefont {Lewis},
  \citenamefont {Marrows},\ and\ \citenamefont {Langridge}}]{Lewis2016}%
  \BibitemOpen
  \bibfield  {author} {\bibinfo {author} {\bibfnamefont {L.~H.}\ \bibnamefont
  {Lewis}}, \bibinfo {author} {\bibfnamefont {C.~H.}\ \bibnamefont {Marrows}},
  \ and\ \bibinfo {author} {\bibfnamefont {S.}~\bibnamefont {Langridge}},\
  }\bibfield  {title} {\enquote {\bibinfo {title} {{Coupled magnetic,
  structural, and electronic phase transitions in FeRh}},}\ }\href {\doibase
  10.1088/0022-3727/49/32/323002} {\bibfield  {journal} {\bibinfo  {journal}
  {J. Phys. D. Appl. Phys.}\ }\textbf {\bibinfo {volume} {49}},\ \bibinfo
  {pages} {323002} (\bibinfo {year} {2016})}\BibitemShut {NoStop}%
\bibitem [{\citenamefont {Maat}\ \emph {et~al.}(2005)\citenamefont {Maat},
  \citenamefont {Thiele},\ and\ \citenamefont {Fullerton}}]{Maat2005}%
  \BibitemOpen
  \bibfield  {author} {\bibinfo {author} {\bibfnamefont {S.}~\bibnamefont
  {Maat}}, \bibinfo {author} {\bibfnamefont {J.-U.}\ \bibnamefont {Thiele}}, \
  and\ \bibinfo {author} {\bibfnamefont {Eric~E.}\ \bibnamefont {Fullerton}},\
  }\bibfield  {title} {\enquote {\bibinfo {title} {{Temperature and field
  hysteresis of the antiferromagnetic-to-ferromagnetic phase transition in
  epitaxial FeRh films}},}\ }\href {\doibase 10.1103/PhysRevB.72.214432}
  {\bibfield  {journal} {\bibinfo  {journal} {Phys. Rev. B}\ }\textbf {\bibinfo
  {volume} {72}},\ \bibinfo {pages} {214432} (\bibinfo {year}
  {2005})}\BibitemShut {NoStop}%
\bibitem [{\citenamefont {Liu}\ \emph {et~al.}(2016)\citenamefont {Liu},
  \citenamefont {Li}, \citenamefont {Gai}, \citenamefont {Clarkson},
  \citenamefont {Hsu}, \citenamefont {Wong}, \citenamefont {Fan}, \citenamefont
  {Lin}, \citenamefont {Rouleau}, \citenamefont {Ward}, \citenamefont {Lee},
  \citenamefont {Sefat}, \citenamefont {Christen},\ and\ \citenamefont
  {Ramesh}}]{Liu2016a}%
  \BibitemOpen
  \bibfield  {author} {\bibinfo {author} {\bibfnamefont {Z.~Q.}\ \bibnamefont
  {Liu}}, \bibinfo {author} {\bibfnamefont {L.}~\bibnamefont {Li}}, \bibinfo
  {author} {\bibfnamefont {Z.}~\bibnamefont {Gai}}, \bibinfo {author}
  {\bibfnamefont {J.~D.}\ \bibnamefont {Clarkson}}, \bibinfo {author}
  {\bibfnamefont {S.~L.}\ \bibnamefont {Hsu}}, \bibinfo {author} {\bibfnamefont
  {A.~T.}\ \bibnamefont {Wong}}, \bibinfo {author} {\bibfnamefont {L.~S.}\
  \bibnamefont {Fan}}, \bibinfo {author} {\bibfnamefont {M.~W.}\ \bibnamefont
  {Lin}}, \bibinfo {author} {\bibfnamefont {C.~M.}\ \bibnamefont {Rouleau}},
  \bibinfo {author} {\bibfnamefont {T.~Z.}\ \bibnamefont {Ward}}, \bibinfo
  {author} {\bibfnamefont {H.~N.}\ \bibnamefont {Lee}}, \bibinfo {author}
  {\bibfnamefont {A.~S.}\ \bibnamefont {Sefat}}, \bibinfo {author}
  {\bibfnamefont {H.~M.}\ \bibnamefont {Christen}}, \ and\ \bibinfo {author}
  {\bibfnamefont {R.}~\bibnamefont {Ramesh}},\ }\bibfield  {title} {\enquote
  {\bibinfo {title} {{Full Electroresistance Modulation in a Mixed-Phase
  Metallic Alloy}},}\ }\href {\doibase 10.1103/PhysRevLett.116.097203}
  {\bibfield  {journal} {\bibinfo  {journal} {Phys. Rev. Lett.}\ }\textbf
  {\bibinfo {volume} {116}},\ \bibinfo {pages} {097203} (\bibinfo {year}
  {2016})}\BibitemShut {NoStop}%
\bibitem [{\citenamefont {Cherifi}\ \emph {et~al.}(2014)\citenamefont
  {Cherifi}, \citenamefont {Ivanovskaya}, \citenamefont {Phillips},
  \citenamefont {Zobelli}, \citenamefont {Infante}, \citenamefont {Jacquet},
  \citenamefont {Garcia}, \citenamefont {Fusil}, \citenamefont {Briddon},
  \citenamefont {Guiblin}, \citenamefont {Mougin}, \citenamefont {{\"{U}}nal},
  \citenamefont {Kronast}, \citenamefont {Valencia}, \citenamefont {Dkhil},
  \citenamefont {Barth{\'{e}}l{\'{e}}my},\ and\ \citenamefont
  {Bibes}}]{Cherifi2014}%
  \BibitemOpen
  \bibfield  {author} {\bibinfo {author} {\bibfnamefont {R.~O.}\ \bibnamefont
  {Cherifi}}, \bibinfo {author} {\bibfnamefont {V.}~\bibnamefont
  {Ivanovskaya}}, \bibinfo {author} {\bibfnamefont {L.~C.}\ \bibnamefont
  {Phillips}}, \bibinfo {author} {\bibfnamefont {A.}~\bibnamefont {Zobelli}},
  \bibinfo {author} {\bibfnamefont {I.~C.}\ \bibnamefont {Infante}}, \bibinfo
  {author} {\bibfnamefont {E.}~\bibnamefont {Jacquet}}, \bibinfo {author}
  {\bibfnamefont {V.}~\bibnamefont {Garcia}}, \bibinfo {author} {\bibfnamefont
  {S.}~\bibnamefont {Fusil}}, \bibinfo {author} {\bibfnamefont {P.~R.}\
  \bibnamefont {Briddon}}, \bibinfo {author} {\bibfnamefont {N.}~\bibnamefont
  {Guiblin}}, \bibinfo {author} {\bibfnamefont {A.}~\bibnamefont {Mougin}},
  \bibinfo {author} {\bibfnamefont {A.~A.}\ \bibnamefont {{\"{U}}nal}},
  \bibinfo {author} {\bibfnamefont {F.}~\bibnamefont {Kronast}}, \bibinfo
  {author} {\bibfnamefont {S.}~\bibnamefont {Valencia}}, \bibinfo {author}
  {\bibfnamefont {B.}~\bibnamefont {Dkhil}}, \bibinfo {author} {\bibfnamefont
  {A.}~\bibnamefont {Barth{\'{e}}l{\'{e}}my}}, \ and\ \bibinfo {author}
  {\bibfnamefont {M.}~\bibnamefont {Bibes}},\ }\bibfield  {title} {\enquote
  {\bibinfo {title} {{Electric-field control of magnetic order above room
  temperature}},}\ }\href {\doibase 10.1038/nmat3870} {\bibfield  {journal}
  {\bibinfo  {journal} {Nat. Mater.}\ }\textbf {\bibinfo {volume} {13}},\
  \bibinfo {pages} {345--351} (\bibinfo {year} {2014})}\BibitemShut {NoStop}%
\bibitem [{\citenamefont {Barua}\ \emph {et~al.}(2013)\citenamefont {Barua},
  \citenamefont {Jim\'{e}nez-Villacorta},\ and\ \citenamefont
  {Lewis}}]{Barua2013b}%
  \BibitemOpen
  \bibfield  {author} {\bibinfo {author} {\bibfnamefont {Radhika}\ \bibnamefont
  {Barua}}, \bibinfo {author} {\bibfnamefont {F\`{e}lix}\ \bibnamefont
  {Jim\'{e}nez-Villacorta}}, \ and\ \bibinfo {author} {\bibfnamefont {L.~H.}\
  \bibnamefont {Lewis}},\ }\bibfield  {title} {\enquote {\bibinfo {title}
  {{Predicting magnetostructural trends in FeRh-based ternary systems}},}\
  }\href {\doibase 10.1063/1.4820583} {\bibfield  {journal} {\bibinfo
  {journal} {Appl. Phys. Lett.}\ }\textbf {\bibinfo {volume} {103}},\ \bibinfo
  {pages} {102407} (\bibinfo {year} {2013})}\BibitemShut {NoStop}%
\bibitem [{\citenamefont {Suzuki}\ \emph {et~al.}(2015)\citenamefont {Suzuki},
  \citenamefont {Naito}, \citenamefont {Itoh},\ and\ \citenamefont
  {Taniyama}}]{Suzuki2015}%
  \BibitemOpen
  \bibfield  {author} {\bibinfo {author} {\bibfnamefont {Ippei}\ \bibnamefont
  {Suzuki}}, \bibinfo {author} {\bibfnamefont {Tomoyuki}\ \bibnamefont
  {Naito}}, \bibinfo {author} {\bibfnamefont {Mitsuru}\ \bibnamefont {Itoh}}, \
  and\ \bibinfo {author} {\bibfnamefont {Tomoyasu}\ \bibnamefont {Taniyama}},\
  }\bibfield  {title} {\enquote {\bibinfo {title} {{Barkhausen-like
  antiferromagnetic to ferromagnetic phase transition driven by spin polarized
  current}},}\ }\href {\doibase 10.1063/1.4929695} {\bibfield  {journal}
  {\bibinfo  {journal} {Appl. Phys. Lett.}\ }\textbf {\bibinfo {volume}
  {107}},\ \bibinfo {pages} {082408} (\bibinfo {year} {2015})}\BibitemShut
  {NoStop}%
\bibitem [{\citenamefont {Thiele}\ \emph {et~al.}(2003)\citenamefont {Thiele},
  \citenamefont {Maat},\ and\ \citenamefont {Fullerton}}]{Thiele2003}%
  \BibitemOpen
  \bibfield  {author} {\bibinfo {author} {\bibfnamefont {Jan-Ulrich}\
  \bibnamefont {Thiele}}, \bibinfo {author} {\bibfnamefont {Stefan}\
  \bibnamefont {Maat}}, \ and\ \bibinfo {author} {\bibfnamefont {Eric~E.}\
  \bibnamefont {Fullerton}},\ }\bibfield  {title} {\enquote {\bibinfo {title}
  {{FeRh/FePt exchange spring films for thermally assisted magnetic recording
  media}},}\ }\href {\doibase 10.1063/1.1571232} {\bibfield  {journal}
  {\bibinfo  {journal} {Appl. Phys. Lett.}\ }\textbf {\bibinfo {volume} {82}},\
  \bibinfo {pages} {2859} (\bibinfo {year} {2003})}\BibitemShut {NoStop}%
\bibitem [{\citenamefont {Marti}\ \emph {et~al.}(2014)\citenamefont {Marti},
  \citenamefont {Fina}, \citenamefont {Frontera}, \citenamefont {Liu},
  \citenamefont {Wadley}, \citenamefont {He}, \citenamefont {Paull},
  \citenamefont {Clarkson}, \citenamefont {Kudrnovsk{\'{y}}}, \citenamefont
  {Turek}, \citenamefont {Kune{\v{s}}}, \citenamefont {Yi}, \citenamefont
  {Chu}, \citenamefont {Nelson}, \citenamefont {You}, \citenamefont {Arenholz},
  \citenamefont {Salahuddin}, \citenamefont {Fontcuberta}, \citenamefont
  {Jungwirth},\ and\ \citenamefont {Ramesh}}]{Marti2014}%
  \BibitemOpen
  \bibfield  {author} {\bibinfo {author} {\bibfnamefont {X.}~\bibnamefont
  {Marti}}, \bibinfo {author} {\bibfnamefont {I.}~\bibnamefont {Fina}},
  \bibinfo {author} {\bibfnamefont {C.}~\bibnamefont {Frontera}}, \bibinfo
  {author} {\bibfnamefont {Jian}\ \bibnamefont {Liu}}, \bibinfo {author}
  {\bibfnamefont {P.}~\bibnamefont {Wadley}}, \bibinfo {author} {\bibfnamefont
  {Q.}~\bibnamefont {He}}, \bibinfo {author} {\bibfnamefont {R.~J.}\
  \bibnamefont {Paull}}, \bibinfo {author} {\bibfnamefont {J.~D.}\ \bibnamefont
  {Clarkson}}, \bibinfo {author} {\bibfnamefont {J.}~\bibnamefont
  {Kudrnovsk{\'{y}}}}, \bibinfo {author} {\bibfnamefont {I.}~\bibnamefont
  {Turek}}, \bibinfo {author} {\bibfnamefont {J.}~\bibnamefont {Kune{\v{s}}}},
  \bibinfo {author} {\bibfnamefont {D.}~\bibnamefont {Yi}}, \bibinfo {author}
  {\bibfnamefont {J-H.}\ \bibnamefont {Chu}}, \bibinfo {author} {\bibfnamefont
  {C.~T.}\ \bibnamefont {Nelson}}, \bibinfo {author} {\bibfnamefont
  {L.}~\bibnamefont {You}}, \bibinfo {author} {\bibfnamefont {E.}~\bibnamefont
  {Arenholz}}, \bibinfo {author} {\bibfnamefont {S.}~\bibnamefont
  {Salahuddin}}, \bibinfo {author} {\bibfnamefont {J.}~\bibnamefont
  {Fontcuberta}}, \bibinfo {author} {\bibfnamefont {T.}~\bibnamefont
  {Jungwirth}}, \ and\ \bibinfo {author} {\bibfnamefont {R.}~\bibnamefont
  {Ramesh}},\ }\bibfield  {title} {\enquote {\bibinfo {title}
  {{Room-temperature antiferromagnetic memory resistor}},}\ }\href {\doibase
  10.1038/nmat3861} {\bibfield  {journal} {\bibinfo  {journal} {Nat. Mater.}\
  }\textbf {\bibinfo {volume} {13}},\ \bibinfo {pages} {367--374} (\bibinfo
  {year} {2014})}\BibitemShut {NoStop}%
\bibitem [{\citenamefont {Moriyama}\ \emph {et~al.}(2015)\citenamefont
  {Moriyama}, \citenamefont {Matsuzaki}, \citenamefont {Kim}, \citenamefont
  {Suzuki}, \citenamefont {Taniyama},\ and\ \citenamefont
  {Ono}}]{Moriyama2015}%
  \BibitemOpen
  \bibfield  {author} {\bibinfo {author} {\bibfnamefont {Takahiro}\
  \bibnamefont {Moriyama}}, \bibinfo {author} {\bibfnamefont {Noriko}\
  \bibnamefont {Matsuzaki}}, \bibinfo {author} {\bibfnamefont {Kab~Jin}\
  \bibnamefont {Kim}}, \bibinfo {author} {\bibfnamefont {Ippei}\ \bibnamefont
  {Suzuki}}, \bibinfo {author} {\bibfnamefont {Tomoyasu}\ \bibnamefont
  {Taniyama}}, \ and\ \bibinfo {author} {\bibfnamefont {Teruo}\ \bibnamefont
  {Ono}},\ }\bibfield  {title} {\enquote {\bibinfo {title} {{Sequential
  write-read operations in FeRh antiferromagnetic memory}},}\ }\href {\doibase
  10.1063/1.4931567} {\bibfield  {journal} {\bibinfo  {journal} {Appl. Phys.
  Lett.}\ }\textbf {\bibinfo {volume} {107}},\ \bibinfo {pages} {122403}
  (\bibinfo {year} {2015})}\BibitemShut {NoStop}%
\bibitem [{\citenamefont {Baldasseroni}\ \emph {et~al.}(2012)\citenamefont
  {Baldasseroni}, \citenamefont {Bordel}, \citenamefont {Gray}, \citenamefont
  {Kaiser}, \citenamefont {Kronast}, \citenamefont {Herrero-Albillos},
  \citenamefont {Schneider}, \citenamefont {Fadley},\ and\ \citenamefont
  {Hellman}}]{Baldasseroni2012}%
  \BibitemOpen
  \bibfield  {author} {\bibinfo {author} {\bibfnamefont {C.}~\bibnamefont
  {Baldasseroni}}, \bibinfo {author} {\bibfnamefont {C.}~\bibnamefont
  {Bordel}}, \bibinfo {author} {\bibfnamefont {A.~X.}\ \bibnamefont {Gray}},
  \bibinfo {author} {\bibfnamefont {A.~M.}\ \bibnamefont {Kaiser}}, \bibinfo
  {author} {\bibfnamefont {F.}~\bibnamefont {Kronast}}, \bibinfo {author}
  {\bibfnamefont {J.}~\bibnamefont {Herrero-Albillos}}, \bibinfo {author}
  {\bibfnamefont {C.~M.}\ \bibnamefont {Schneider}}, \bibinfo {author}
  {\bibfnamefont {C.~S.}\ \bibnamefont {Fadley}}, \ and\ \bibinfo {author}
  {\bibfnamefont {F.}~\bibnamefont {Hellman}},\ }\bibfield  {title} {\enquote
  {\bibinfo {title} {{Temperature-driven nucleation of ferromagnetic domains in
  FeRh thin films}},}\ }\href {\doibase 10.1063/1.4730957} {\bibfield
  {journal} {\bibinfo  {journal} {Appl. Phys. Lett.}\ }\textbf {\bibinfo
  {volume} {100}},\ \bibinfo {pages} {262401} (\bibinfo {year}
  {2012})}\BibitemShut {NoStop}%
\bibitem [{\citenamefont {Kinane}\ \emph {et~al.}(2014)\citenamefont {Kinane},
  \citenamefont {Loving}, \citenamefont {de~Vries}, \citenamefont {Fan},
  \citenamefont {Charlton}, \citenamefont {Claydon}, \citenamefont {Arena},
  \citenamefont {Maccherozzi}, \citenamefont {Dhesi}, \citenamefont {Heiman},
  \citenamefont {Marrows}, \citenamefont {Lewis},\ and\ \citenamefont
  {Langridge}}]{Kinane2014}%
  \BibitemOpen
  \bibfield  {author} {\bibinfo {author} {\bibfnamefont {C~J}\ \bibnamefont
  {Kinane}}, \bibinfo {author} {\bibfnamefont {M}~\bibnamefont {Loving}},
  \bibinfo {author} {\bibfnamefont {M~A}\ \bibnamefont {de~Vries}}, \bibinfo
  {author} {\bibfnamefont {R}~\bibnamefont {Fan}}, \bibinfo {author}
  {\bibfnamefont {T~R}\ \bibnamefont {Charlton}}, \bibinfo {author}
  {\bibfnamefont {J~S}\ \bibnamefont {Claydon}}, \bibinfo {author}
  {\bibfnamefont {D~A}\ \bibnamefont {Arena}}, \bibinfo {author} {\bibfnamefont
  {F}~\bibnamefont {Maccherozzi}}, \bibinfo {author} {\bibfnamefont {S~S}\
  \bibnamefont {Dhesi}}, \bibinfo {author} {\bibfnamefont {D}~\bibnamefont
  {Heiman}}, \bibinfo {author} {\bibfnamefont {C~H}\ \bibnamefont {Marrows}},
  \bibinfo {author} {\bibfnamefont {L~H}\ \bibnamefont {Lewis}}, \ and\
  \bibinfo {author} {\bibfnamefont {Sean}\ \bibnamefont {Langridge}},\
  }\bibfield  {title} {\enquote {\bibinfo {title} {{Observation of a
  temperature dependent asymmetry in the domain structure of a Pd-doped FeRh
  epilayer}},}\ }\href {\doibase 10.1088/1367-2630/16/11/113073} {\bibfield
  {journal} {\bibinfo  {journal} {New J. Phys.}\ }\textbf {\bibinfo {volume}
  {16}},\ \bibinfo {pages} {113073} (\bibinfo {year} {2014})}\BibitemShut
  {NoStop}%
\bibitem [{\citenamefont {Baldasseroni}\ \emph {et~al.}(2015)\citenamefont
  {Baldasseroni}, \citenamefont {Bordel}, \citenamefont {Antonakos},
  \citenamefont {Scholl}, \citenamefont {Stone}, \citenamefont {Kortright},\
  and\ \citenamefont {Hellman}}]{Baldasseroni2015}%
  \BibitemOpen
  \bibfield  {author} {\bibinfo {author} {\bibfnamefont {C}~\bibnamefont
  {Baldasseroni}}, \bibinfo {author} {\bibfnamefont {C}~\bibnamefont {Bordel}},
  \bibinfo {author} {\bibfnamefont {C}~\bibnamefont {Antonakos}}, \bibinfo
  {author} {\bibfnamefont {a}~\bibnamefont {Scholl}}, \bibinfo {author}
  {\bibfnamefont {K~H}\ \bibnamefont {Stone}}, \bibinfo {author} {\bibfnamefont
  {J~B}\ \bibnamefont {Kortright}}, \ and\ \bibinfo {author} {\bibfnamefont
  {F}~\bibnamefont {Hellman}},\ }\bibfield  {title} {\enquote {\bibinfo {title}
  {{Temperature-driven growth of antiferromagnetic domains in thin-film
  FeRh}},}\ }\href {\doibase 10.1088/0953-8984/27/25/256001} {\bibfield
  {journal} {\bibinfo  {journal} {J. Phys. Condens. Matter}\ }\textbf {\bibinfo
  {volume} {27}},\ \bibinfo {pages} {256001} (\bibinfo {year}
  {2015})}\BibitemShut {NoStop}%
\bibitem [{\citenamefont {Almeida}\ \emph {et~al.}(2017)\citenamefont
  {Almeida}, \citenamefont {Temple}, \citenamefont {Massey}, \citenamefont
  {Fallon}, \citenamefont {McGrouther}, \citenamefont {Moore}, \citenamefont
  {Marrows},\ and\ \citenamefont {McVitie}}]{Almeida2017a}%
  \BibitemOpen
  \bibfield  {author} {\bibinfo {author} {\bibfnamefont {Trevor~P.}\
  \bibnamefont {Almeida}}, \bibinfo {author} {\bibfnamefont {Rowan}\
  \bibnamefont {Temple}}, \bibinfo {author} {\bibfnamefont {Jamie}\
  \bibnamefont {Massey}}, \bibinfo {author} {\bibfnamefont {Kayla}\
  \bibnamefont {Fallon}}, \bibinfo {author} {\bibfnamefont {Damien}\
  \bibnamefont {McGrouther}}, \bibinfo {author} {\bibfnamefont {Thomas}\
  \bibnamefont {Moore}}, \bibinfo {author} {\bibfnamefont {Christopher~H.}\
  \bibnamefont {Marrows}}, \ and\ \bibinfo {author} {\bibfnamefont {Stephen}\
  \bibnamefont {McVitie}},\ }\bibfield  {title} {\enquote {\bibinfo {title}
  {{Quantitative TEM imaging of the magnetostructural and phase transitions in
  FeRh thin film systems}},}\ }\href {\doibase 10.1038/s41598-017-18194-0}
  {\bibfield  {journal} {\bibinfo  {journal} {Sci. Rep.}\ }\textbf {\bibinfo
  {volume} {7}},\ \bibinfo {pages} {17835} (\bibinfo {year}
  {2017})}\BibitemShut {NoStop}%
\bibitem [{\citenamefont {{Le Gra{\"{e}}t}}\ \emph {et~al.}(2015)\citenamefont
  {{Le Gra{\"{e}}t}}, \citenamefont {Charlton}, \citenamefont {McLaren},
  \citenamefont {Loving}, \citenamefont {Morley}, \citenamefont {Kinane},
  \citenamefont {Brydson}, \citenamefont {Lewis}, \citenamefont {Langridge},\
  and\ \citenamefont {Marrows}}]{LeGraet2015}%
  \BibitemOpen
  \bibfield  {author} {\bibinfo {author} {\bibfnamefont {C.}~\bibnamefont {{Le
  Gra{\"{e}}t}}}, \bibinfo {author} {\bibfnamefont {T.~R.}\ \bibnamefont
  {Charlton}}, \bibinfo {author} {\bibfnamefont {M.}~\bibnamefont {McLaren}},
  \bibinfo {author} {\bibfnamefont {M.}~\bibnamefont {Loving}}, \bibinfo
  {author} {\bibfnamefont {S.~A.}\ \bibnamefont {Morley}}, \bibinfo {author}
  {\bibfnamefont {C.~J.}\ \bibnamefont {Kinane}}, \bibinfo {author}
  {\bibfnamefont {R.~M.~D.}\ \bibnamefont {Brydson}}, \bibinfo {author}
  {\bibfnamefont {L.~H.}\ \bibnamefont {Lewis}}, \bibinfo {author}
  {\bibfnamefont {S.}~\bibnamefont {Langridge}}, \ and\ \bibinfo {author}
  {\bibfnamefont {C.~H.}\ \bibnamefont {Marrows}},\ }\bibfield  {title}
  {\enquote {\bibinfo {title} {{Temperature controlled motion of an
  antiferromagnet- ferromagnet interface within a dopant-graded FeRh
  epilayer}},}\ }\href {\doibase 10.1063/1.4907282} {\bibfield  {journal}
  {\bibinfo  {journal} {APL Mater.}\ }\textbf {\bibinfo {volume} {3}},\
  \bibinfo {pages} {041802} (\bibinfo {year} {2015})},\ \Eprint
  {http://arxiv.org/abs/1412.7346} {arXiv:1412.7346} \BibitemShut {NoStop}%
\bibitem [{\citenamefont {Ceballos}\ \emph {et~al.}(2017)\citenamefont
  {Ceballos}, \citenamefont {Chen}, \citenamefont {Schneider}, \citenamefont
  {Bordel}, \citenamefont {Wang},\ and\ \citenamefont
  {Hellman}}]{Ceballos2017}%
  \BibitemOpen
  \bibfield  {author} {\bibinfo {author} {\bibfnamefont {A.}~\bibnamefont
  {Ceballos}}, \bibinfo {author} {\bibfnamefont {Zhanghui}\ \bibnamefont
  {Chen}}, \bibinfo {author} {\bibfnamefont {O.}~\bibnamefont {Schneider}},
  \bibinfo {author} {\bibfnamefont {C.}~\bibnamefont {Bordel}}, \bibinfo
  {author} {\bibfnamefont {Lin~Wang}\ \bibnamefont {Wang}}, \ and\ \bibinfo
  {author} {\bibfnamefont {F.}~\bibnamefont {Hellman}},\ }\bibfield  {title}
  {\enquote {\bibinfo {title} {{Effect of strain and thickness on the
  transition temperature of epitaxial FeRh thin-films}},}\ }\href {\doibase
  10.1063/1.4997901} {\bibfield  {journal} {\bibinfo  {journal} {Appl. Phys.
  Lett.}\ }\textbf {\bibinfo {volume} {111}},\ \bibinfo {pages} {172401}
  (\bibinfo {year} {2017})}\BibitemShut {NoStop}%
\bibitem [{\citenamefont {Gatel}\ \emph {et~al.}(2017)\citenamefont {Gatel},
  \citenamefont {Warot-Fonrose}, \citenamefont {Biziere}, \citenamefont
  {Rodr{\'{i}}guez}, \citenamefont {Reyes}, \citenamefont {Cours},
  \citenamefont {Castiella},\ and\ \citenamefont {Casanove}}]{Gatel2017}%
  \BibitemOpen
  \bibfield  {author} {\bibinfo {author} {\bibfnamefont {C.}~\bibnamefont
  {Gatel}}, \bibinfo {author} {\bibfnamefont {B.}~\bibnamefont
  {Warot-Fonrose}}, \bibinfo {author} {\bibfnamefont {N.}~\bibnamefont
  {Biziere}}, \bibinfo {author} {\bibfnamefont {L.~A.}\ \bibnamefont
  {Rodr{\'{i}}guez}}, \bibinfo {author} {\bibfnamefont {D.}~\bibnamefont
  {Reyes}}, \bibinfo {author} {\bibfnamefont {R.}~\bibnamefont {Cours}},
  \bibinfo {author} {\bibfnamefont {M.}~\bibnamefont {Castiella}}, \ and\
  \bibinfo {author} {\bibfnamefont {M.~J.}\ \bibnamefont {Casanove}},\
  }\bibfield  {title} {\enquote {\bibinfo {title} {{Inhomogeneous spatial
  distribution of the magnetic transition in an iron-rhodium thin film}},}\
  }\href {\doibase 10.1038/ncomms15703} {\bibfield  {journal} {\bibinfo
  {journal} {Nat. Commun.}\ }\textbf {\bibinfo {volume} {8}},\ \bibinfo {pages}
  {15703} (\bibinfo {year} {2017})}\BibitemShut {NoStop}%
\bibitem [{\citenamefont {Pressacco}\ \emph {et~al.}(2016)\citenamefont
  {Pressacco}, \citenamefont {Uhl{\'{i}}ř}, \citenamefont {Gatti},
  \citenamefont {Bendounan}, \citenamefont {Fullerton},\ and\ \citenamefont
  {Sirotti}}]{Pressacco2016}%
  \BibitemOpen
  \bibfield  {author} {\bibinfo {author} {\bibfnamefont {Federico}\
  \bibnamefont {Pressacco}}, \bibinfo {author} {\bibfnamefont {Vojt{\v{e}}ch}\
  \bibnamefont {Uhl{\'{i}}ř}}, \bibinfo {author} {\bibfnamefont {Matteo}\
  \bibnamefont {Gatti}}, \bibinfo {author} {\bibfnamefont {Azzedine}\
  \bibnamefont {Bendounan}}, \bibinfo {author} {\bibfnamefont {Eric~E.}\
  \bibnamefont {Fullerton}}, \ and\ \bibinfo {author} {\bibfnamefont {Fausto}\
  \bibnamefont {Sirotti}},\ }\bibfield  {title} {\enquote {\bibinfo {title}
  {{Stable room-temperature ferromagnetic phase at the FeRh(100) surface}},}\
  }\href {\doibase 10.1038/srep22383} {\bibfield  {journal} {\bibinfo
  {journal} {Sci. Rep.}\ }\textbf {\bibinfo {volume} {6}},\ \bibinfo {pages}
  {22383} (\bibinfo {year} {2016})},\ \Eprint {http://arxiv.org/abs/1508.01777}
  {arXiv:1508.01777} \BibitemShut {NoStop}%
\bibitem [{\citenamefont {Barton}\ \emph {et~al.}(2017)\citenamefont {Barton},
  \citenamefont {Ostler}, \citenamefont {Huskisson}, \citenamefont {Kinane},
  \citenamefont {Haigh}, \citenamefont {Hrkac},\ and\ \citenamefont
  {Thomson}}]{Barton2017}%
  \BibitemOpen
  \bibfield  {author} {\bibinfo {author} {\bibfnamefont {C.~W.}\ \bibnamefont
  {Barton}}, \bibinfo {author} {\bibfnamefont {T.~A.}\ \bibnamefont {Ostler}},
  \bibinfo {author} {\bibfnamefont {D.}~\bibnamefont {Huskisson}}, \bibinfo
  {author} {\bibfnamefont {C.~J.}\ \bibnamefont {Kinane}}, \bibinfo {author}
  {\bibfnamefont {S.~J.}\ \bibnamefont {Haigh}}, \bibinfo {author}
  {\bibfnamefont {G.}~\bibnamefont {Hrkac}}, \ and\ \bibinfo {author}
  {\bibfnamefont {T.}~\bibnamefont {Thomson}},\ }\bibfield  {title} {\enquote
  {\bibinfo {title} {{Substrate Induced Strain Field in FeRh Epilayers Grown on
  Single Crystal MgO (001) Substrates}},}\ }\href {\doibase 10.1038/srep44397}
  {\bibfield  {journal} {\bibinfo  {journal} {Sci. Rep.}\ }\textbf {\bibinfo
  {volume} {7}},\ \bibinfo {pages} {44397} (\bibinfo {year}
  {2017})}\BibitemShut {NoStop}%
\bibitem [{\citenamefont {Uhl{\'{i}}ř}\ \emph {et~al.}(2016)\citenamefont
  {Uhl{\'{i}}ř}, \citenamefont {Arregi},\ and\ \citenamefont
  {Fullerton}}]{Uhlir2016}%
  \BibitemOpen
  \bibfield  {author} {\bibinfo {author} {\bibfnamefont {V.}~\bibnamefont
  {Uhl{\'{i}}ř}}, \bibinfo {author} {\bibfnamefont {J.~A.}\ \bibnamefont
  {Arregi}}, \ and\ \bibinfo {author} {\bibfnamefont {E.~E.}\ \bibnamefont
  {Fullerton}},\ }\bibfield  {title} {\enquote {\bibinfo {title} {{Colossal
  magnetic phase transition asymmetry in mesoscale FeRh stripes}},}\ }\href
  {\doibase 10.1038/ncomms13113} {\bibfield  {journal} {\bibinfo  {journal}
  {Nat. Commun.}\ }\textbf {\bibinfo {volume} {7}},\ \bibinfo {pages} {13113}
  (\bibinfo {year} {2016})},\ \Eprint {http://arxiv.org/abs/1605.06823}
  {arXiv:1605.06823} \BibitemShut {NoStop}%
\bibitem [{\citenamefont {Arregi}\ \emph {et~al.}(2018)\citenamefont {Arregi},
  \citenamefont {Hork{\'{y}}}, \citenamefont {Fabianov{\'{a}}}, \citenamefont
  {Tolley}, \citenamefont {Fullerton},\ and\ \citenamefont
  {Uhliř}}]{Arregi2018}%
  \BibitemOpen
  \bibfield  {author} {\bibinfo {author} {\bibfnamefont {Jon~Ander}\
  \bibnamefont {Arregi}}, \bibinfo {author} {\bibfnamefont {Michal}\
  \bibnamefont {Hork{\'{y}}}}, \bibinfo {author} {\bibfnamefont {Kateřina}\
  \bibnamefont {Fabianov{\'{a}}}}, \bibinfo {author} {\bibfnamefont {Robert}\
  \bibnamefont {Tolley}}, \bibinfo {author} {\bibfnamefont {Eric~E.}\
  \bibnamefont {Fullerton}}, \ and\ \bibinfo {author} {\bibfnamefont
  {Vojt{\v{e}}ch}\ \bibnamefont {Uhliř}},\ }\bibfield  {title} {\enquote
  {\bibinfo {title} {{Magnetization reversal and confinement effects across the
  metamagnetic phase transition in mesoscale FeRh structures}},}\ }\href
  {\doibase 10.1088/1361-6463/aaaa5a} {\bibfield  {journal} {\bibinfo
  {journal} {J. Phys. D. Appl. Phys.}\ }\textbf {\bibinfo {volume} {51}},\
  \bibinfo {pages} {105001} (\bibinfo {year} {2018})}\BibitemShut {NoStop}%
\bibitem [{\citenamefont {{Le Gra{\"{e}}t}}\ \emph {et~al.}(2013)\citenamefont
  {{Le Gra{\"{e}}t}}, \citenamefont {de~Vries}, \citenamefont {McLaren},
  \citenamefont {Brydson}, \citenamefont {Loving}, \citenamefont {Heiman},
  \citenamefont {Lewis},\ and\ \citenamefont {Marrows}}]{LeGraet2013}%
  \BibitemOpen
  \bibfield  {author} {\bibinfo {author} {\bibfnamefont {Chantal}\ \bibnamefont
  {{Le Gra{\"{e}}t}}}, \bibinfo {author} {\bibfnamefont {Mark~A.}\ \bibnamefont
  {de~Vries}}, \bibinfo {author} {\bibfnamefont {Mathew}\ \bibnamefont
  {McLaren}}, \bibinfo {author} {\bibfnamefont {Richard~M.D.}\ \bibnamefont
  {Brydson}}, \bibinfo {author} {\bibfnamefont {Melissa}\ \bibnamefont
  {Loving}}, \bibinfo {author} {\bibfnamefont {Don}\ \bibnamefont {Heiman}},
  \bibinfo {author} {\bibfnamefont {Laura~H.}\ \bibnamefont {Lewis}}, \ and\
  \bibinfo {author} {\bibfnamefont {Christopher~H.}\ \bibnamefont {Marrows}},\
  }\bibfield  {title} {\enquote {\bibinfo {title} {{Sputter Growth and
  Characterization of Metamagnetic B2-ordered FeRh Epilayers}},}\ }\href
  {\doibase 10.3791/50603} {\bibfield  {journal} {\bibinfo  {journal} {J. Vis.
  Exp.}\ ,\ \bibinfo {pages} {e50603}} (\bibinfo {year} {2013})}\BibitemShut
  {NoStop}%
\bibitem [{\citenamefont {Baldasseroni}\ \emph {et~al.}(2014)\citenamefont
  {Baldasseroni}, \citenamefont {P{\'{a}}lsson}, \citenamefont {Bordel},
  \citenamefont {Valencia}, \citenamefont {Unal}, \citenamefont {Kronast},
  \citenamefont {Nemsak}, \citenamefont {Fadley}, \citenamefont {Borchers},
  \citenamefont {Maranville},\ and\ \citenamefont
  {Hellman}}]{Baldasseroni2014}%
  \BibitemOpen
  \bibfield  {author} {\bibinfo {author} {\bibfnamefont {C.}~\bibnamefont
  {Baldasseroni}}, \bibinfo {author} {\bibfnamefont {G.~K.}\ \bibnamefont
  {P{\'{a}}lsson}}, \bibinfo {author} {\bibfnamefont {C.}~\bibnamefont
  {Bordel}}, \bibinfo {author} {\bibfnamefont {S.}~\bibnamefont {Valencia}},
  \bibinfo {author} {\bibfnamefont {a.~a.}\ \bibnamefont {Unal}}, \bibinfo
  {author} {\bibfnamefont {F.}~\bibnamefont {Kronast}}, \bibinfo {author}
  {\bibfnamefont {S.}~\bibnamefont {Nemsak}}, \bibinfo {author} {\bibfnamefont
  {C.~S.}\ \bibnamefont {Fadley}}, \bibinfo {author} {\bibfnamefont {J.~a.}\
  \bibnamefont {Borchers}}, \bibinfo {author} {\bibfnamefont {B.~B.}\
  \bibnamefont {Maranville}}, \ and\ \bibinfo {author} {\bibfnamefont
  {F.}~\bibnamefont {Hellman}},\ }\bibfield  {title} {\enquote {\bibinfo
  {title} {{Effect of capping material on interfacial ferromagnetism in FeRh
  thin films}},}\ }\href {\doibase 10.1063/1.4862961} {\bibfield  {journal}
  {\bibinfo  {journal} {J. Appl. Phys.}\ }\textbf {\bibinfo {volume} {115}},\
  \bibinfo {pages} {043919} (\bibinfo {year} {2014})}\BibitemShut {NoStop}%
\bibitem [{\citenamefont {de~Vries}\ \emph {et~al.}(2013)\citenamefont
  {de~Vries}, \citenamefont {Loving}, \citenamefont {Mihai}, \citenamefont
  {Lewis}, \citenamefont {Heiman},\ and\ \citenamefont
  {Marrows}}]{deVries2013}%
  \BibitemOpen
  \bibfield  {author} {\bibinfo {author} {\bibfnamefont {M~A}\ \bibnamefont
  {de~Vries}}, \bibinfo {author} {\bibfnamefont {M}~\bibnamefont {Loving}},
  \bibinfo {author} {\bibfnamefont {A~P}\ \bibnamefont {Mihai}}, \bibinfo
  {author} {\bibfnamefont {L~H}\ \bibnamefont {Lewis}}, \bibinfo {author}
  {\bibfnamefont {D}~\bibnamefont {Heiman}}, \ and\ \bibinfo {author}
  {\bibfnamefont {C~H}\ \bibnamefont {Marrows}},\ }\bibfield  {title} {\enquote
  {\bibinfo {title} {{Hall-effect characterization of the metamagnetic
  transition in FeRh}},}\ }\href {\doibase 10.1088/1367-2630/15/1/013008}
  {\bibfield  {journal} {\bibinfo  {journal} {New J. Phys.}\ }\textbf {\bibinfo
  {volume} {15}},\ \bibinfo {pages} {013008} (\bibinfo {year}
  {2013})}\BibitemShut {NoStop}%
\bibitem [{\citenamefont {Kim}\ \emph {et~al.}(2009)\citenamefont {Kim},
  \citenamefont {Ryan}, \citenamefont {Ding}, \citenamefont {Lewis},
  \citenamefont {Ali}, \citenamefont {Kinane}, \citenamefont {Hickey},
  \citenamefont {Marrows},\ and\ \citenamefont {Arena}}]{Kim2009}%
  \BibitemOpen
  \bibfield  {author} {\bibinfo {author} {\bibfnamefont {J.~W.}\ \bibnamefont
  {Kim}}, \bibinfo {author} {\bibfnamefont {P.~J.}\ \bibnamefont {Ryan}},
  \bibinfo {author} {\bibfnamefont {Y.}~\bibnamefont {Ding}}, \bibinfo {author}
  {\bibfnamefont {L.~H.}\ \bibnamefont {Lewis}}, \bibinfo {author}
  {\bibfnamefont {M.}~\bibnamefont {Ali}}, \bibinfo {author} {\bibfnamefont
  {C.~J.}\ \bibnamefont {Kinane}}, \bibinfo {author} {\bibfnamefont {B.~J.}\
  \bibnamefont {Hickey}}, \bibinfo {author} {\bibfnamefont {C.~H.}\
  \bibnamefont {Marrows}}, \ and\ \bibinfo {author} {\bibfnamefont {D.~A.}\
  \bibnamefont {Arena}},\ }\bibfield  {title} {\enquote {\bibinfo {title}
  {{Surface influenced magnetostructural transition in FeRh films}},}\ }\href
  {\doibase 10.1063/1.3265921} {\bibfield  {journal} {\bibinfo  {journal}
  {Appl. Phys. Lett.}\ }\textbf {\bibinfo {volume} {95}},\ \bibinfo {pages}
  {222515} (\bibinfo {year} {2009})}\BibitemShut {NoStop}%
\bibitem [{\citenamefont {Fan}\ \emph {et~al.}(2010)\citenamefont {Fan},
  \citenamefont {Kinane}, \citenamefont {Charlton}, \citenamefont {Dorner},
  \citenamefont {Ali}, \citenamefont {de~Vries}, \citenamefont {Brydson},
  \citenamefont {Marrows}, \citenamefont {Hickey}, \citenamefont {Arena},
  \citenamefont {Tanner}, \citenamefont {Nisbet},\ and\ \citenamefont
  {Langridge}}]{Fan2010}%
  \BibitemOpen
  \bibfield  {author} {\bibinfo {author} {\bibfnamefont {R.}~\bibnamefont
  {Fan}}, \bibinfo {author} {\bibfnamefont {C.~J.}\ \bibnamefont {Kinane}},
  \bibinfo {author} {\bibfnamefont {T.~R.}\ \bibnamefont {Charlton}}, \bibinfo
  {author} {\bibfnamefont {R.}~\bibnamefont {Dorner}}, \bibinfo {author}
  {\bibfnamefont {M.}~\bibnamefont {Ali}}, \bibinfo {author} {\bibfnamefont
  {M.~A.}\ \bibnamefont {de~Vries}}, \bibinfo {author} {\bibfnamefont
  {R.~M.D.}\ \bibnamefont {Brydson}}, \bibinfo {author} {\bibfnamefont {C.~H.}\
  \bibnamefont {Marrows}}, \bibinfo {author} {\bibfnamefont {B.~J.}\
  \bibnamefont {Hickey}}, \bibinfo {author} {\bibfnamefont {D.~A.}\
  \bibnamefont {Arena}}, \bibinfo {author} {\bibfnamefont {B.~K.}\ \bibnamefont
  {Tanner}}, \bibinfo {author} {\bibfnamefont {G.}~\bibnamefont {Nisbet}}, \
  and\ \bibinfo {author} {\bibfnamefont {S.}~\bibnamefont {Langridge}},\
  }\bibfield  {title} {\enquote {\bibinfo {title} {{Ferromagnetism at the
  interfaces of antiferromagnetic FeRh epilayers}},}\ }\href {\doibase
  10.1103/PhysRevB.82.184418} {\bibfield  {journal} {\bibinfo  {journal} {Phys.
  Rev. B - Condens. Matter Mater. Phys.}\ }\textbf {\bibinfo {volume} {82}},\
  \bibinfo {pages} {184418} (\bibinfo {year} {2010})}\BibitemShut {NoStop}%
\bibitem [{\citenamefont {Mariager}\ \emph {et~al.}(2013)\citenamefont
  {Mariager}, \citenamefont {Guyader}, \citenamefont {Buzzi}, \citenamefont
  {Ingold},\ and\ \citenamefont {Quitmann}}]{Mariager2013}%
  \BibitemOpen
  \bibfield  {author} {\bibinfo {author} {\bibfnamefont {S.~O.}\ \bibnamefont
  {Mariager}}, \bibinfo {author} {\bibfnamefont {L.~Le}\ \bibnamefont
  {Guyader}}, \bibinfo {author} {\bibfnamefont {M}~\bibnamefont {Buzzi}},
  \bibinfo {author} {\bibfnamefont {G.}~\bibnamefont {Ingold}}, \ and\ \bibinfo
  {author} {\bibfnamefont {C.}~\bibnamefont {Quitmann}},\ }\href
  {http://arxiv.org/abs/1301.4164} {\enquote {\bibinfo {title} {{Imaging the
  antiferromagnetic to ferromagnetic first order phase transition of FeRh}},}\
  } (\bibinfo {year} {2013}),\ \Eprint {http://arxiv.org/abs/1301.4164}
  {arXiv:1301.4164} \BibitemShut {NoStop}%
\bibitem [{\citenamefont {Cohen}(1962)}]{Cohen1962}%
  \BibitemOpen
  \bibfield  {author} {\bibinfo {author} {\bibfnamefont {M~S}\ \bibnamefont
  {Cohen}},\ }\bibfield  {title} {\enquote {\bibinfo {title} {{Anomalous
  Magnetic Films}},}\ }\href {\doibase 10.1063/1.1728545} {\bibfield  {journal}
  {\bibinfo  {journal} {J. Appl. Phys.}\ }\textbf {\bibinfo {volume} {33}},\
  \bibinfo {pages} {2968--2980} (\bibinfo {year} {1962})}\BibitemShut {NoStop}%
\bibitem [{\citenamefont {Himcinschi}\ \emph {et~al.}(2007)\citenamefont
  {Himcinschi}, \citenamefont {Singh}, \citenamefont {Radu}, \citenamefont
  {Milenin}, \citenamefont {Erfurth}, \citenamefont {Reiche}, \citenamefont
  {G{\"{o}}sele}, \citenamefont {Christiansen}, \citenamefont {Muster},\ and\
  \citenamefont {Petzold}}]{Himcinschi2007}%
  \BibitemOpen
  \bibfield  {author} {\bibinfo {author} {\bibfnamefont {C.}~\bibnamefont
  {Himcinschi}}, \bibinfo {author} {\bibfnamefont {R.}~\bibnamefont {Singh}},
  \bibinfo {author} {\bibfnamefont {I.}~\bibnamefont {Radu}}, \bibinfo {author}
  {\bibfnamefont {A.~P.}\ \bibnamefont {Milenin}}, \bibinfo {author}
  {\bibfnamefont {W.}~\bibnamefont {Erfurth}}, \bibinfo {author} {\bibfnamefont
  {M.}~\bibnamefont {Reiche}}, \bibinfo {author} {\bibfnamefont
  {U.}~\bibnamefont {G{\"{o}}sele}}, \bibinfo {author} {\bibfnamefont {S.~H.}\
  \bibnamefont {Christiansen}}, \bibinfo {author} {\bibfnamefont
  {F.}~\bibnamefont {Muster}}, \ and\ \bibinfo {author} {\bibfnamefont
  {M.}~\bibnamefont {Petzold}},\ }\bibfield  {title} {\enquote {\bibinfo
  {title} {{Strain relaxation in nanopatterned strained silicon round
  pillars}},}\ }\href {\doibase 10.1063/1.2431476} {\bibfield  {journal}
  {\bibinfo  {journal} {Appl. Phys. Lett.}\ }\textbf {\bibinfo {volume} {90}},\
  \bibinfo {pages} {021902} (\bibinfo {year} {2007})}\BibitemShut {NoStop}%
\bibitem [{\citenamefont {Shaw}\ \emph {et~al.}(2008)\citenamefont {Shaw},
  \citenamefont {Russek}, \citenamefont {Thomson}, \citenamefont {Donahue},
  \citenamefont {Terris}, \citenamefont {Hellwig}, \citenamefont {Dobisz},\
  and\ \citenamefont {Schneider}}]{Shaw2008}%
  \BibitemOpen
  \bibfield  {author} {\bibinfo {author} {\bibfnamefont {Justin~M.}\
  \bibnamefont {Shaw}}, \bibinfo {author} {\bibfnamefont {Stephen~E.}\
  \bibnamefont {Russek}}, \bibinfo {author} {\bibfnamefont {Thomas}\
  \bibnamefont {Thomson}}, \bibinfo {author} {\bibfnamefont {Michael~J.}\
  \bibnamefont {Donahue}}, \bibinfo {author} {\bibfnamefont {Bruce~D.}\
  \bibnamefont {Terris}}, \bibinfo {author} {\bibfnamefont {Olav}\ \bibnamefont
  {Hellwig}}, \bibinfo {author} {\bibfnamefont {Elizabeth}\ \bibnamefont
  {Dobisz}}, \ and\ \bibinfo {author} {\bibfnamefont {Michael~L.}\ \bibnamefont
  {Schneider}},\ }\bibfield  {title} {\enquote {\bibinfo {title} {{Reversal
  mechanisms in perpendicularly magnetized nanostructures}},}\ }\href {\doibase
  10.1103/PhysRevB.78.024414} {\bibfield  {journal} {\bibinfo  {journal} {Phys.
  Rev. B - Condens. Matter Mater. Phys.}\ }\textbf {\bibinfo {volume} {78}},\
  \bibinfo {pages} {024414} (\bibinfo {year} {2008})}\BibitemShut {NoStop}%
\bibitem [{\citenamefont {Cervera}\ \emph {et~al.}(2017)\citenamefont
  {Cervera}, \citenamefont {Trassinelli}, \citenamefont {Marangolo},
  \citenamefont {Carr{\'{e}}t{\'{e}}ro}, \citenamefont {Garcia}, \citenamefont
  {Hidki}, \citenamefont {Jacquet}, \citenamefont {Lamour}, \citenamefont
  {L{\'{e}}vy}, \citenamefont {Mac{\'{e}}}, \citenamefont {Prigent},
  \citenamefont {Rozet}, \citenamefont {Steydli},\ and\ \citenamefont
  {Vernhet}}]{Cervera2017}%
  \BibitemOpen
  \bibfield  {author} {\bibinfo {author} {\bibfnamefont {S.}~\bibnamefont
  {Cervera}}, \bibinfo {author} {\bibfnamefont {M}~\bibnamefont {Trassinelli}},
  \bibinfo {author} {\bibfnamefont {M}~\bibnamefont {Marangolo}}, \bibinfo
  {author} {\bibfnamefont {C}~\bibnamefont {Carr{\'{e}}t{\'{e}}ro}}, \bibinfo
  {author} {\bibfnamefont {V}~\bibnamefont {Garcia}}, \bibinfo {author}
  {\bibfnamefont {S}~\bibnamefont {Hidki}}, \bibinfo {author} {\bibfnamefont
  {E}~\bibnamefont {Jacquet}}, \bibinfo {author} {\bibfnamefont
  {E}~\bibnamefont {Lamour}}, \bibinfo {author} {\bibfnamefont {A}~\bibnamefont
  {L{\'{e}}vy}}, \bibinfo {author} {\bibfnamefont {S}~\bibnamefont
  {Mac{\'{e}}}}, \bibinfo {author} {\bibfnamefont {C}~\bibnamefont {Prigent}},
  \bibinfo {author} {\bibfnamefont {J}~\bibnamefont {Rozet}}, \bibinfo {author}
  {\bibfnamefont {S}~\bibnamefont {Steydli}}, \ and\ \bibinfo {author}
  {\bibfnamefont {D}~\bibnamefont {Vernhet}},\ }\bibfield  {title} {\enquote
  {\bibinfo {title} {{Modulating the phase transition temperature of giant
  magnetocaloric thin films by ion irradiation}},}\ }\href {\doibase
  10.1103/PhysRevMaterials.1.065402} {\bibfield  {journal} {\bibinfo  {journal}
  {Phys. Rev. Mat.}\ }\textbf {\bibinfo {volume} {1}},\ \bibinfo {pages}
  {065402} (\bibinfo {year} {2017})},\ \Eprint
  {http://arxiv.org/abs/1710.08229} {arXiv:1710.08229} \BibitemShut {NoStop}%
\bibitem [{\citenamefont {Mclaren}(2014)}]{McLarenThesis}%
  \BibitemOpen
  \bibfield  {author} {\bibinfo {author} {\bibfnamefont {M.~J.}\ \bibnamefont
  {Mclaren}},\ }\emph {\bibinfo {title} {Transmission electron microscope
  characterisation of iron-rhodium epilayers}},\ \href
  {http://etheses.whiterose.ac.uk/} {Ph.D. thesis} (\bibinfo {year}
  {2014})\BibitemShut {NoStop}%
\bibitem [{\citenamefont {Zeigler}()}]{TRIMsoftware}%
  \BibitemOpen
  \bibfield  {author} {\bibinfo {author} {\bibfnamefont {J.~F.}\ \bibnamefont
  {Zeigler}},\ }\href@noop {} {\enquote {\bibinfo {title} {{SRIM and TRIM}},}\
  }\bibinfo {note} {{http://www.srim.org/}}\BibitemShut {NoStop}%
\bibitem [{\citenamefont {Dobrev}(1982)}]{Dobrev1982}%
  \BibitemOpen
  \bibfield  {author} {\bibinfo {author} {\bibfnamefont {D.}~\bibnamefont
  {Dobrev}},\ }\bibfield  {title} {\enquote {\bibinfo {title} {{Ion-beam
  induced texture formation in vacuum condensed thin metal films}},}\
  }\href@noop {} {\bibfield  {journal} {\bibinfo  {journal} {Thin Solid Films}\
  }\textbf {\bibinfo {volume} {92}},\ \bibinfo {pages} {41--53} (\bibinfo
  {year} {1982})}\BibitemShut {NoStop}%
\bibitem [{Qua()}]{QuantumDetectors}%
  \BibitemOpen
  \href@noop {} {\enquote {\bibinfo {title} {{Merlin for EM}},}\ }\bibinfo
  {note} {{http://quantumdetectors.com/merlin-for-em/}}\BibitemShut {NoStop}%
\end{thebibliography}
